\documentclass[]{article}

\usepackage{epsfig}
\usepackage{amsmath}
\usepackage{amsthm}
\usepackage{amssymb}
\usepackage{longtable}

\DeclareGraphicsExtensions{.pdf,.png}
\graphicspath{{imgs/}}

\providecommand{\idtt}[1]{\ensuremath{\mathtt{#1}}}
\newcommand{\rank}{\idtt{rank}}
\newcommand{\select}{\idtt{select}}
\newcommand{\access}{\idtt{access}}
\newcommand{\map}{\idtt{map}}
\newcommand{\unmap}{\idtt{unmap}}
\newcommand{\lab}{\idtt{lab}}
\newcommand{\poslab}{\idtt{poslab}}

\newcommand{\R}{\mathcal{R}}
\newcommand{\La}{\mathcal{L}}

\newcommand{\relrnk}{\idtt{rel\_rnk}}

\newcommand{\relnum}{\idtt{rel\_num}}
\newcommand{\relselobjfst}{\idtt{rel\_sel\_obj\_maj}}
\newcommand{\relsellabfst}{\idtt{rel\_sel\_lab\_maj}}
\newcommand{\relminobjfst}{\idtt{rel\_min\_obj\_maj}}
\newcommand{\relminlabfst}{\idtt{rel\_min\_lab\_maj}}
\newcommand{\relacc}{\idtt{rel\_acc}}
\newcommand{\relaccobjfst}{\idtt{rel\_acc\_obj\_maj}}
\newcommand{\relacclabfst}{\idtt{rel\_acc\_lab\_maj}}
\newcommand{\relrnkobjfst}{\idtt{rel\_rnk\_obj\_maj}}
\newcommand{\relrnklabfst}{\idtt{rel\_rnk\_lab\_maj}}
\newcommand{\labsel}{\idtt{lab\_sel}}
\newcommand{\objsel}{\idtt{obj\_sel}}
\newcommand{\labacc}{\idtt{lab\_acc}}
\newcommand{\objacc}{\idtt{obj\_acc}}
\newcommand{\labmin}{\idtt{lab\_min}}
\newcommand{\objmin}{\idtt{obj\_min}}
\newcommand{\objrnk}{\idtt{obj\_rnk}}
\newcommand{\labrnk}{\idtt{lab\_rnk}}
\newcommand{\objnum}{\idtt{obj\_num}}
\newcommand{\labnum}{\idtt{lab\_num}}
\newcommand{\objrnko}{\idtt{obj\_rnk1}}
\newcommand{\labrnko}{\idtt{lab\_rnk1}}
\newcommand{\objselo}{\idtt{obj\_sel1}}
\newcommand{\labselo}{\idtt{lab\_sel1}}
\newcommand{\objacco}{\idtt{obj\_acc1}}
\newcommand{\labacco}{\idtt{lab\_acc1}}
\newcommand{\objmino}{\idtt{obj\_min1}}
\newcommand{\labmino}{\idtt{lab\_min1}}

\newcommand{\child}{\idtt{child}}
\newcommand{\g}{\idtt{g}}

\renewcommand{\paragraph}{\noindent$\bullet$ \textbf}

\newcommand{\no}[1]{}

\newtheorem{theorem}{Theorem}
\newtheorem{lemma}{Lemma}

\begin{document}

\title{Compact Binary Relation Representations \\ with Rich Functionality
\thanks{An early version of this article appeared in {\em Proc. LATIN 2010}.Partially funded by Fondecyt grant 1-110066, Chile (first and third authors). Funded in part by Google U.S./Canada PhD Fellowship Program and David R. Cheriton Scholarships Program (second author).}}

\author{J\'er\'emy Barbay \\ University of Chile \\ \text{jbarbay@dcc.uchile.cl}
\and
Francisco Claude \\ University of Waterloo \\ \text{fclaude@cs.uwaterloo.ca}
\and
Gonzalo Navarro \\ University of Chile \\ \text{gnavarro@dcc.uchile.cl}}

\maketitle

\begin{abstract}
  Binary relations are an important abstraction arising in many 
  data representation problems.  
  The data structures proposed so far to represent them support just a few 
  basic operations required to fit one particular application.
  We identify many of those operations arising in applications and generalize
  them into a wide set of desirable queries for a binary relation
  representation. We also identify reductions among those operations.
  We then introduce several novel binary relation representations, some simple 
  and some quite sophisticated, that not only are space-efficient but also 
  efficiently support a large subset of the desired queries.
\end{abstract}

\section{Introduction}

Binary relations appear everywhere in Computer Science. Graphs, trees,
inverted indexes, strings and permutations are just some examples.  They
also arise as a tool to complement existing data structures (such
as trees~\cite{BGMR07} or graphs~\cite{BCAHM07}) with additional
information, such as weights or labels on the nodes or edges, that can be
indexed and searched.  Interestingly, the data structure support for binary
relations has not undergone a systematic study, but rather one triggered by
particular applications. We aim to start such a study in this article.

\begin{figure}[t]
\begin{center}
\includegraphics[width=4.5cm]{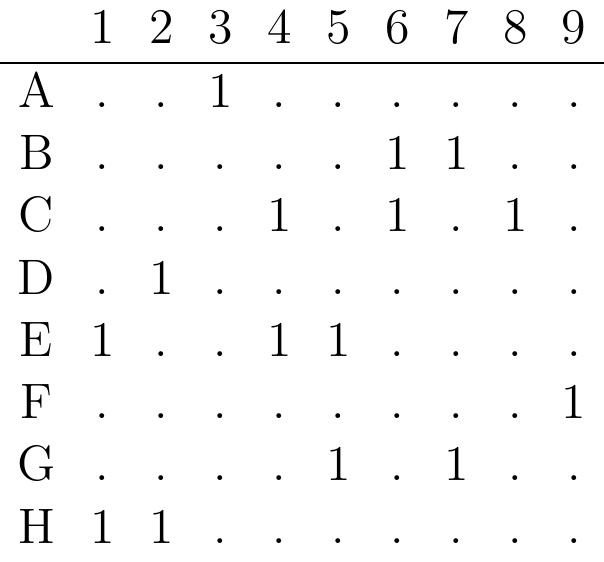}
\caption{An example of binary relation.}
\label{fig:binrel_example}
\end{center}
\end{figure}

Let us say that a binary relation $\R$ relates {\em objects} in
$[1,n]$ with {\em labels} in $[1,\sigma]$, containing $t$ pairs out of the
$n\sigma$ possible ones. We focus on space-efficient representations
considering a simple entropy measure,
\[ 
   H(\R) ~~=~~ \lg {n\sigma \choose t} ~~=~~ t\lg\frac{n\sigma}{t} + O(t)
\]
bits, which ignores any other possible regularity.
Figure~\ref{fig:binrel_example} illustrates a binary relation (we identify
labels with rows and objects with columns henceforth).

Previous work focused on relatively basic primitives for binary relations:
extract the list of all the labels associated with an object or of all the 
objects associated with a label (an operation called $\access$), or extracting 
the $j$-th such element (an operation called $\select$), or counting how many
of these are there up to some object/label value (called operation
$\rank$). 

The first representation specifically designed for binary
relations~\cite{BGMR07} supports $\rank$, $\select$ and $\access$ on the
rows (labels) of the relation, for the purpose of supporting faster joins
on labels.  The idea is to write the labels of the pairs in object-major
order and to operate on the resulting string plus some auxiliary data. This
approach was extended to support more general operations needed for text
indexing \cite{CNfi10}.
The first technique~\cite{BGMR07} was later refined~\cite{BHMR07} into a scheme
that allows one to compress the string while still supporting the basic
operations on both labels and objects. The idea is to store auxiliary data
on top of an arbitrary representation of the binary relation, which can
thus be compressed. This was used to support labeled operations on planar
and quasi-planar labeled graphs~\cite{BCAHM07}.

Ad-hoc compressed representations for inverted lists~\cite{WMB99} and Web
graphs~\cite{CNtweb10} can also be considered as supporting binary relations.
The idea here is to write the objects of the pairs, in label-major order,
and to support extracting substrings of the resulting string, that is,
little more than $\access$ on labels. One can add support for $\access$ on
objects by means of string $\select$ operations \cite{CNlncs10}.
The string can be compressed by different means depending on the application.

In this paper we aim at settling the foundations of efficient compact
data structures for binary relations. In particular, we address the
following points:

\begin{itemize}
\item We define a large set of operations of relevance to binary relations,
widely extending the classic set of $\rank$, $\select$ and $\access$. We give
a number of reductions among operations in order to define a core set that 
allows one to efficiently support the extended set of operations.
\item We explore the power of the reduction to string
operators~\cite{BGMR07} when the operations supported on the string are
limited to $\rank$, $\select$ and $\access$. This structure is called
{\sc BinRel-Str}, and we show it achieves
interesting bounds only for a reduced set of operations.
\item We show that a particular string representation, the wavelet 
tree~\cite{GGV03}, although not being the fastest one, provides native support 
for a much wider set of operations within logarithmic time. We call {\sc
BinRel-WT} this binary relation representation.
\item We extend wavelet trees to generalized wavelet trees~\cite{FMMN07}, 
and design new algorithms for various operations that take advantage of their
larger fan-out. As a result we speed up most of the operations within the same
space. This structure is called {\sc BinRel-GWT}.
\item We present a new structure, {\em binary relation wavelet tree}
({\sc BRWT}), that is tailored to represent binary relations. Although the
{\sc BRWT} gives weaker support to the operations, it is the only one that
approaches the entropy space $H(\R)$ within a multiplicative factor 
(of 1.272).
\end{itemize}

For the sake of brevity, we aim at the simplest description of the
operations, ignoring any practical improvement that does not make a
difference in terms of asymptotic time complexity, or trivial extensions 
such as interchanging labels and objects to obtain other space/time tradeoffs.

\section{Compact Data Structures for Sequences}
\label{sec:seqs}

Given a sequence $S$ of length $n$, drawn from an alphabet $\Sigma$ of size
$\sigma$, we define the following queries (omitting $S$ if clear from 
context):

\begin{itemize}
\item $\rank_a(S,i)$ counts the occurrences of symbol $a\in\Sigma$ in
$S[1,i]$.
\item $\select_a(S,j)$ finds the position of the $j$-th occurrence of symbol $a\in\Sigma$
in $S$.
\item $\access(S,i)=S[i]$. 
\end{itemize}

The solutions in the literature are quite different depending on the alphabet 
size, $\sigma$.

\subsection{Binary Sequences}

For the special case $\Sigma=\{0,1\}$, there exist representations using
$o(n)$ bits on top of a plain representation of $S$, and answering the three
queries in constant time \cite{Cla96}. The extra space can be made as low as
$O(n\lg\lg n / \lg n)$, which is optimal \cite{Gol07}. This extra data
structure on top of a plain representation of $S$ is called an {\em index}.

If we are allowed, instead, to represent $S$ in a specific way, then one can
{\em compress} the sequence while still supporting the three operations in
constant time. The overall space can be reduced to $nH_0(S)+o(n)$ bits
\cite{RRR02}. Here $H_0(S)$ is the {\em zero-order entropy} of sequence
$S$, defined as 
\[ H_0(S) ~~=~~ \sum_{a\in\Sigma} \frac{n_a}{n}\lg\frac{n}{n_a},
\]
where $n_a$ is the number of occurrences of symbol $a$ in $S$. The $o(n)$
extra space on top of the entropy can be made as small as $O(n/\lg^c n)$ for
any constant $c$ \cite{Pat08}.

\subsection{Sequences over Small Alphabets}
\label{sec:seqsmall}

If $\sigma = O(\lg^\epsilon n)$, for a constant $0<\epsilon<1$, it is still
possible to retain the space and time complexities of binary sequence
representations \cite{FMMN07}. The main idea is that we only need a compressed
sequence representation that provides constant-time $\access$, or more
precisely, that can access $O(\lg_\sigma n)$ consecutive symbols from $S$ in
constant time (this is achieved by extending similar compressed bitmap 
representations \cite{RRR02}).

Then, in order to support $\rank_a$ and $\select_a$, we act as if we had an
explicit bitmap $B_a[1,n]$, where $B_a[i] = 1$ iff $S[i]=a$. Then 
$\rank_a(S,i) = \rank_1(B_a,i)$ and $\select_a(S,j)=\select_1(B_a,j)$.
We store only the $\rank$/$\select$ {\em index} of each $B_a$, and can solve
$\rank$/$\select$ on $B_a$ by extracting any desired chunk from $S$. The
only difference is that we cannot access $O(\lg n)$ contiguous bits from $B_a$
in constant time, but only $O(\lg_\sigma n)$.

As a result, the index for each $B_a$ requires $O(n \lg\lg n / \lg_\sigma n)$
bits of space. Added over all the $a \in \Sigma$, the total space for the
indexes is $O(n \sigma \lg\lg n / \lg_\sigma n) = o(n)$.

\subsection{General Sequences and Wavelet Trees}
\label{sec:seqlarge}

The wavelet tree \cite{GGV03} reduces the three operations on general
alphabets to those on binary sequences. It is a perfectly balanced tree
that stores a bitmap of length $n$ at the root; every position in the
bitmap is either 0 or 1 depending on whether the symbol at this position
belongs to the first half of the alphabet or to the second. The left child
of the root will handle the subsequence of $S$ marked with a 0 at the root,
and the right child will handle the 1s. This decomposition into alphabet
subranges continues recursively until reaching level $\lceil \lg \sigma
\rceil$, where the leaves correspond to individual symbols. We call $B_v$
the bitmap at node $v$.

The $\access$ query $S[i]$ can be answered by following the path described
for position $i$. At the root $v$, if $B_v[i] = 0/1$, we descend to the
left/right child, switching to the bitmap position $\rank_{0/1}(B_v,i)$ in
the left/right child, which then becomes the new $v$.  This continues
recursively until reaching the last level, when we arrive at the leaf
corresponding to the desired symbol.
Query $\rank_a(S,i)$ can be answered similarly to $\access$, except that we
descend according to $a$ and not to the bit of $B_v$.  We update position
$i$ for the child node just as before.  At the leaves, the final bitmap
position $i$ is the answer.
Query $\select_a(S,j)$ proceeds as $\rank_a$, but upwards. We start at the
leaf representing $a$ and update $j$ to $\select_{0/1}(B_v,j)$ where $v$ is
the parent node, depending on whether the current node is its left/right
child. At the root, position $j$ is the result.

\begin{figure}[t]
\begin{center}
\includegraphics[width=11cm]{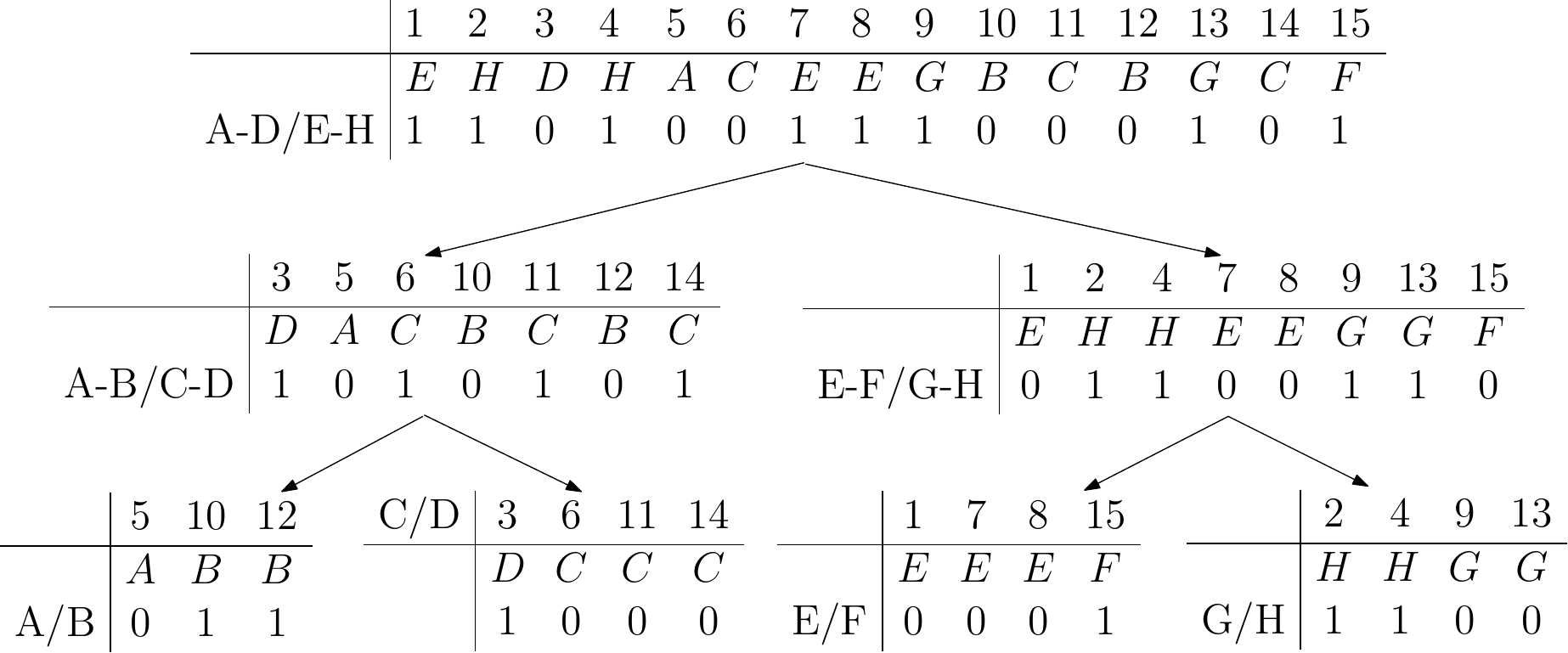}
\caption{Example of a wavelet tree for the sequence {\tt EHDHACEEGBCBGCF}.}
\label{fig:wt_example}
\end{center}
\end{figure}

If the bitmaps $B_v$ are represented in plain form (with indexes that support
binary $\rank$/$\select$), then the wavelet tree requires 
$n\lg\sigma+o(n)\lg\sigma$ bits of space, while answering all the queries in 
$O(\lg\sigma)$ time. If the bitmaps $B_v$ are instead represented in compressed
form \cite{Pat08}, the wavelet tree uses $nH_0(S) + o(n)$ bits and retains the
same time complexities. 
Figure \ref{fig:wt_example} illustrates the structure. Wavelet 
trees are not only used to represent strings \cite{FMMN07}, but also grids 
\cite{BHMM09}, permutations \cite{BN09}, and many other structures. We refer
in particular to a recent article \cite{GNP11} where in particular some
(folklore) capabilities are carefully proved: (1) any range of symbols
$[\alpha,\beta]$ is covered by $k=O(\lg(\beta-\alpha+1))$ wavelet tree nodes,
and these can be reached from the root by traversing $O(k+\lg\sigma)$ nodes.

A way to speed up the wavelet tree operations is to use {\em generalized}
wavelet trees \cite{FMMN07}. These are multiary wavelet trees, with arity
$\mu = O(\lg^\epsilon n)$, for a constant $0<\epsilon<1$. Bitmaps $B_v$ are
replaced by sequences $S_v$ over a (small) alphabet of size $\mu$. All the
operations on those sequences are still solved in constant time, but now the
wavelet tree height is reduced to $O(\lg_\mu \sigma) = O(\lg\sigma/\lg\lg n)$,
and thus this is the complexity of the operations. The space can still be
bounded by $nH_0(S)+o(n)$ \cite{GRR08}.

Wavelet trees are not the only sequence representation achieving basically
$nH_0(S)$ bits of space \cite{GMR06,BGNN10}. The best current alternative
\cite{BGNN10} uses $nH_0(S) + o(nH_0(S)) + o(n)$ bits and solves the three
queries within time $O(\lg\lg\sigma)$. This is preferable when $\sigma$ is
large.

\subsection{Range Minimum Queries (RMQs)}
\label{sec:rmq}

The Range Minimum Query (RMQ) operation on a sequence $S[1,n]$ was not listed
among the basic ones, but we cover it because we will make use of it in
Lemma~\ref{lem:gwtop3.5}. It is defined as 
$\textsc{rmq}(S,i,j) = \textrm{argmin}_{i \le k \le j} S[k]$,
that is, it returns the position of the minimum value in a range $S[i,j]$.
If there are more than one, it returns the leftmost minimum.

It is possible to solve this query in constant time using just $2n+o(n)$ bits
of space, without even accessing $S$ itself \cite{Fis10}. 

\section{Operations}

\subsection{Definition of operations}
\label{sec:defin-oper}

We now motivate the set of operations we define for binary relations. Their
full list, formal definition, and illustration, are given in \ref{sec:defop}.

One of the most pervasive examples of binary relations are directed graphs,
which are precisely binary relations between a vertex set $V$ and itself.
Extracting rows or columns in this binary relation supports direct and reverse 
navigation from a node. To support powerful direct access to rows and columns
we define operations $\objacco(\alpha,x,y)$, which retrieves the objects in 
$[x,y]$ related to label $\alpha$, in arbitrary order, and the symmetric one, 
$\labacco(\alpha,\beta,x)$. In case we want to retrieve the pairs in order,
$\objmino(\alpha,x)$, which gives the first object $\ge x$ related to label
$\alpha$, can be used as an iterator, and similarly $\labmino(\alpha,x)$.
These operations are also useful to find out whether the link $(\alpha,x)$
exists. Note that adjacency list representations only support efficiently the
retrieval of direct neighbors, and adjacency matrix only support efficiently
the test for the existence of a link.

Web graphs, and their compact representation supporting navigation, have been 
a subject of intense research in recent years \cite{BV03,BLN09,CNtweb10} (see
many more references therein). In a Web graph, the nodes are Web pages and the 
edges are hyperlinks. Nodes are usually sorted by URL, which not only gives 
good compression but also makes ranges of nodes correspond to domains and 
subdirectories\footnote{More precisely, we have to sort by the reversed site 
string concatenated with the path.}. For example, counting the number of 
connections between two ranges of nodes allows estimating the connectivity 
between two domains. This count of points in a range is supported by our 
operation $\relnum(\alpha,\beta,x,y)$, which counts the number of related pairs
in $[\alpha,\beta] \times [x,y]$. The individual links between the two domains 
can be retrieved, in arbitrary order, with operation
$\relacc(\alpha,\beta,x,y)$. In general, considering domain ranges enables 
the analysis and navigation of the Web graph at a coarser granularity (e.g., 
as a graph of hosts, or institutions). Our operations 
$\objacc(\alpha,\beta,x,y)$, which gives the objects in $[x,y]$ related to
a label in $[\alpha,\beta]$, extends $\objacco$ to ranges of labels,
and similarly $\labacc(\alpha,\beta,x,y)$ extends $\labacco$. Ordered
enumeration and (coarse) link testing are supported by operations
$\objmin(\alpha,\beta,x)$, which gives the first object $\ge x$ related to
a label in $[\alpha,\beta]$, and similarly $\labmin(\alpha,x,y)$ for labels.

A second pervasive example of a binary relation is formed by the two-dimensional
grids, where objects and labels are simply coordinates, and where pairs of the 
relation are points at those coordinates. Grids arise in GIS and
many other geometric applications. Operation $\relnum$ allows us to count
the number of points in a rectangular area. A second essential operation in
these applications is to retrieve the points from such an area. If the retrieval
order is not important, $\relacc$ is sufficient. Otherwise, operation
$\relacclabfst(x,y,\alpha,z)$ serves as an iterator to retrieve the points
in label-major order. It retrieves the first point in $[\alpha,\alpha] \times
[z,y]$, if any, and otherwise the first point in $[\alpha{+}1,\sigma] \times
[x,y]$. Operation $\relaccobjfst(\alpha,\beta,\gamma,x)$ is similar, for
object-major order. For an even more sophisticated processing of the
points, $\relsellabfst(\alpha,j,x,y)$ and $\relselobjfst(\alpha,\beta,x,j)$
give access to the $j$-th element in such lists.

Grids also arise in more abstract scenarios. For example, several text indexing
data structures \cite{CHSV08,CNfi10,Kar99,MNtcs07} resort to a grid, which 
relates for example text suffixes (in lexicographic order) with their text 
positions, or phrase prefixes and suffixes in Lempel-Ziv compression, or two 
labels that form a rule in grammar-based compression, etc.  The operations most 
commonly needed are, again, counting and retrieving (in arbitrary order) the
points in a rectangle.

Another important example of binary relations are inverted indexes~\cite{WMB99},
which support word-based searches on natural language text collections.
Inverted indexes can be seen as a relation between vocabulary words (the 
labels) and the documents where they appear (the objects). Apart from the basic
operation of extracting the documents where a word appears ($\objacco$), a
popular operation is the conjunctive query (e.g., in Google-like search 
engines), which retrieves the documents where $k$ given words appear. These 
are solved using a combination of the complementary queries 
$\objrnko(\alpha,x)$ and $\objselo(\alpha,x,j)$ \cite{BGMR07,BLOLS09}. The first
operation counts the number of points in $[\alpha,\alpha] \times [1,x]$, 
whereas the second gives the $j$-th point in $[\alpha,\alpha] \times [x,n]$.

Extending these operations to a range of words allows for stemmed and/or prefix
searches (by properly ordering the words), and are implemented using
$\objrnk(\alpha,\beta,x)$ and $\objsel(\alpha,\beta,x,j)$, which extend
$\objrnko$ and $\objselo$ to ranges of labels. Extracting a column, on the
other hand ($\labacco$), gives important {\em summarization} information on a 
document: the list of its different words. Intersecting columns (using the
symmetric operations $\labrnko(\alpha,x)$ and $\labselo(\alpha,x,j)$) allows 
for analysis of content between documents (e.g., plagiarism or common 
authorship detection). Handling ranges of documents (supported with the
symmetric operations $\labacc$, $\labrnk(\alpha,x,y)$, and 
$\labsel(\alpha,j,x,y)$) allows for considering hierarchical document
structures such as XML or file systems (where one operates over a whole subtree 
or subdirectory).

Similar representations are useful to support join operations on relational 
databases and, in combination with data structures for ordinal trees, to
support multi-labeled trees, such as those featured by semi-structured
documents (e.g., XML) \cite{BGMR07}.
A similar technique~\cite{BCAHM07} combining various data structures for
graphs with binary relations yields a family of data structures for
edge-labeled and vertex-labeled graphs that support labeled operations on
the neighborhood of each vertex.
For example, operations $\relminlabfst$ and $\relminobjfst$ support the 
search for the highest neighbor of a point, when the binary relation encodes 
the levels of points in a planar graph representing a topography 
map~\cite{BCAHM07}.

The extension of those operations to the union of labels in a given range
allows them to handle more complex queries, such as conjunctions of
disjunctions. For example, in a relational database, consecutive labels may 
represent a range of parameter values (e.g., people of age between $20$ and 
$40$). 

We define other operations for completeness: $\relrnklabfst$ acts like the
inverse of $\relsellabfst$, and similarly $\relrnkobjfst$; $\labnum$ and
$\objnum$ are more complete versions of $\labrnk$ and $\objrnk$; and
$\relrnk$ is a more basic version of $\relnum$.

\subsection{Reductions among operations}
\label{sec:equiv-betw-oper}

We give a set of reductions among the operations introduced. The results are
sumarized in the following theorem.

\begin{figure}[t]
\begin{center}
\includegraphics[width=11cm]{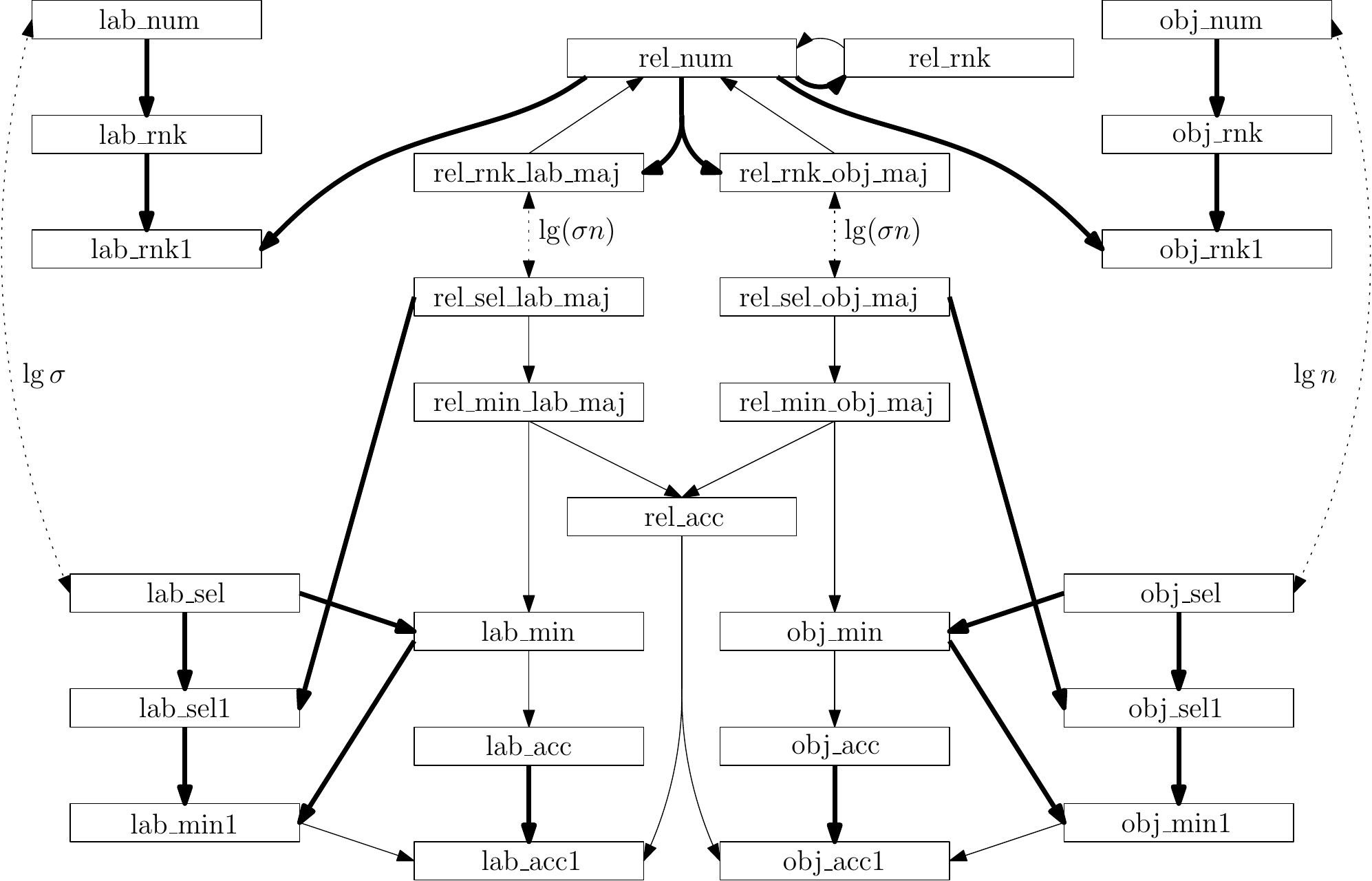}
\caption{Reductions among operations.}
\label{fig:reductions}
\end{center}
\end{figure}

\begin{theorem} \label{thm:reductions}
For any solid arrow $op \rightarrow op'$ in Figure \ref{fig:reductions}, it
holds that if $op$ is solved in time $t$, then $op'$ can be solved in time
$O(t)$. For the dotted arrows with associated penalty factors $O(t')$, it
holds that if $op$ is solved in time $t$, then $op'$ can be solved in time
$O(tt')$.
\end{theorem}

\begin{proof}
Several reductions are immediate from the definition of the operations in 
\ref{sec:defop} (those arrows are in bold in Figure~\ref{fig:reductions}). 
We prove now the other ones. We consider only the reductions 
for the left side of the figure (operations related to labels); the same idea 
applies for the right side (objects).

\begin{itemize}
\item $\relrnk \rightarrow \relnum$
\begin{eqnarray*}
\relnum(\alpha,\beta,x,y) &=& \relrnk(\alpha-1,x-1) - \relrnk(\alpha-1,y) \\
& & - \relrnk(\beta,x-1) + \relrnk(\alpha-1,x-1).
\end{eqnarray*}

\item $\relrnklabfst \rightarrow \relnum$
\begin{eqnarray*}
\relnum(\alpha,\beta,x,y) &=& \relrnklabfst(\beta,x,y,y) \\
& & - \relrnklabfst(\alpha-1,x,y,y).
\end{eqnarray*}

\item $\relacc \rightarrow (\labacco,\objacco)$
\begin{eqnarray*}
\labacco(\alpha,\beta,x) & = & \{ \gamma,\ (\gamma,x) \in \relacc(\alpha,\beta,x,x) \}, \\
\objacco(\alpha,x,y) & = & \{ z,\ (\alpha,z) \in \relacc(\alpha,\alpha,x,y)\}. 
\end{eqnarray*}

\item $\relsellabfst \rightarrow \relminlabfst$: in order to solve query
$\relminlabfst(\alpha,x,y,z)$ we first test if
$\relsellabfst(\alpha,1,z,y)$ gives a pair of the form $(\alpha,w)$, in which
case we return it. Otherwise, we return $\relsellabfst(\alpha+1,1,x,y)$.

\item $\relminlabfst \rightarrow \relacc$: to solve $\relacc(\alpha,\beta,x,y)$,
we find a first point $(\gamma,z) = \relminlabfst(\alpha,x,y,x)$. 
The next element is obtained as $(\gamma',z') = \relminlabfst(\gamma,x,y,z+1)$
and so on, until we reach the first answer with label greater than $\beta$.

\item $\relminlabfst \rightarrow \labmin$: 
let $(\gamma,z) = \relminlabfst(\alpha,x,y,x)$, then
$\labmin(\alpha,x,y) = \gamma$.

\item $\labmin \rightarrow \labacc$: we report $\gamma = \labmin(\alpha,x,y)$,
$\gamma' = \labmin(\gamma+1,x,y)$, and so on until reaching a result larger
than $\beta$. The points reported form $\labacc(\alpha,\beta,x,y)$.

\item $\labmino \rightarrow \labacco$: similar to the previous reduction.
\end{itemize}

Finally, the non-constant time reductions are explained the following way:

\begin{itemize}
\item $(\relrnklabfst,\relsellabfst)$ works in both ways by doing a binary
search over the results of the other operation, in the worst case
considering $n\sigma$ elements.
\item $(\labnum,\labsel)$ operates the same way as the previous one, but
searching among $\sigma$ elements in the worst case, thus the $O(\lg
\sigma)$ penalty. (For the objects this becomes $O(\lg n)$.)
\end{itemize}
\end{proof}

The reductions presented here allow us to focus on a small subset of
the most difficult operations. In some cases, however, we will present
more efficient solutions for the simpler operations and will not use the
reduction.

\section{Reduction to Strings: {\sc BinRel-Str}}
\label{sec:string}

\begin{figure}[t]
\begin{center}
\includegraphics[width=6cm]{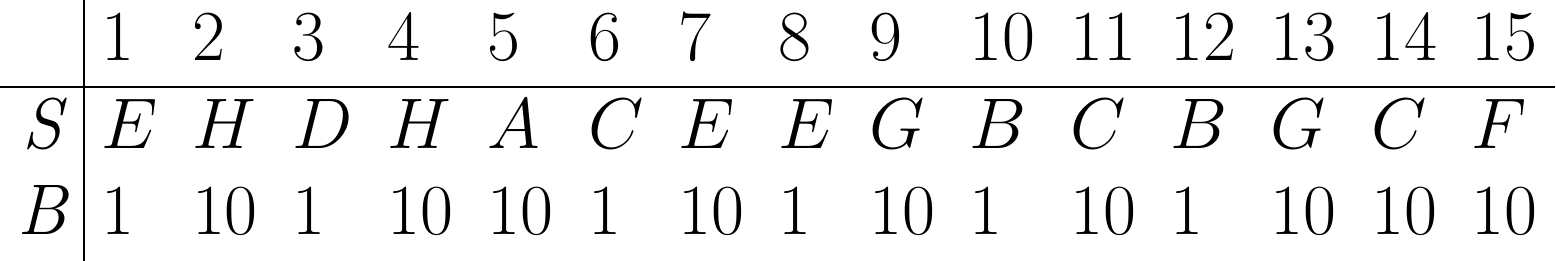}
\caption{Sequence $S$ and bitmap $B$ for representing the binary relation 
shown in Figure~\ref{fig:binrel_example}.}
\label{fig:b_example}
\end{center}
\end{figure}

A simple representation \cite{BGMR07,CNfi10} for a binary relation $\R$ formed 
by $t$ pairs in $[1,n]\times[1,\sigma]$ uses a bitmap $B[1,n+t]$ and a
string $S[1,t]$ over the alphabet $[1,\sigma]$.  The bitmap $B$
concatenates the consecutive cardinalities of the $n$ columns of the
relation, in unary.  The string $S$ contains the rows (labels) of the
pairs of the relation in column (object)-major order. Figure~\ref{fig:b_example} 
shows the representation for the binary
relation shown in Figure~\ref{fig:binrel_example}.  Barbay et
  al.~\cite{BGMR07} showed that an easy way to support operations $\objrnko$
and $\objselo$ on the binary relation is to support the operations $\rank$ and 
$\select$ on $B$ and $S$, using any data structure known for bitmaps and 
strings (recall Section~\ref{sec:seqs}). Note also that the particular case 
$\relnum(1,\sigma,x,y)$ can be answered in $O(1)$ time using $B$. In the sequel
we extend Barbay et al.'s work as much as possible considering our considerably
larger set of operations. This approach, building only on $\rank$, $\select$
and $\access$ on $B$ and $S$, will be called {\sc BinRel-Str}.

We define some notation used in the rest of the paper. First, we call 
$\map(x)$ the mapping from a column number $x$ to its last element in $S$:
$\map(x) = \rank_1(B,\select_0(B,x))$.  The inverse, from a position in $S$ to 
its column number, is called $\unmap(m) = \rank_0(B,\select_1(B,m))+1$. 
Both mappings take constant time.  Finally, let us also define for shortness
$\rank_c(B,x,y) = \rank_c(B,y)-\rank_c(B,x-1)$.

Assume our representation of $S$ supports $\access$ in time $a$, $\rank$ in 
time $r$ and $\select$ in time $s$. Table \ref{tab:time_wt_vs_str} shows the 
complexity achieved for each binary relation operation using this approach. 
As it can be seen, the scheme extends nicely only to operations involving
one row or one column. In all the other cases, the complexities are linear
in the lengths of the ranges to consider. As the algorithms are straightforward
and their complexities uninteresting, we defer them to 
\ref{sec:binrelstr-proofs}.\footnote{To simplify, in 
Table \ref{tab:time_wt_vs_str} we omit some complexities that are most likely 
to be inferior to their alternatives.}

\begin{table}[tbp]
\begin{center}
{\scriptsize 
\begin{tabular}{| l | p{3.3cm} | p{3.3cm} |} \hline
Operation & {\sc BinRel-Str} & {\sc BinRel-WT}\\
\hline\hline
$\relnum(\alpha,\beta,x,y)$ 
& $O((\beta-\alpha+1)r)$ 

  $O((y-x+1)a\lg\beta)$
& $O(\lg\sigma)$ \\ \hline

$\relrnk(\alpha,x)$ 
& 
  $O(\alpha r)$ 

  $O(xa\lg\alpha)$ 
& $O(\lg\sigma)$ \\ \hline

$\relrnklabfst(\alpha,x,y,z)$ 
& $O(\alpha r)$ 

  $O((y-x+1)a\lg\alpha)$  
& $O(\lg\sigma)$\\ \hline

$\relsellabfst(\alpha,j,x,y)$ 
& $O((\sigma-\alpha+1)r+s)$ 
& $O(\lg\sigma)$\\ \hline

$\relminlabfst(\alpha,x,y,z)$ 
& $O((\sigma-\alpha+1)r+s)$ 

  $O((y-x+1)a\lg\alpha)$
& $O(\lg\sigma)$\\ \hline

$\relrnkobjfst(\alpha,\beta,\gamma,x)$ 
&
  $O((\beta-\alpha+1)r)$

  $O(xa\lg\beta)$
& $O(\lg\sigma)$\\ \hline

$\relselobjfst(\alpha,\beta,x,j)$ 
&  
  $O((n-x+1)a\lg\beta)$
& 
  $O(\lg j\lg(\beta-\alpha+1)\lg\sigma)$

  $O(\lg n \lg\sigma)$ \\ \hline

$\relminobjfst(\alpha,\beta,\gamma,x)$ 
& $O((\beta-\alpha+1)(s+r))$

  $O((n-x+1)a\lg\alpha)$
& $O(\lg\sigma)$\\ \hline

$\relacc(\alpha,\beta,x,y)$ 
& $O((\beta-\alpha+1)r+sk)$ 

  $O((y-x+1)a\lg\alpha+ak)$ 
& $O((k+1)\lg\sigma)$\\ \hline

$\labnum(\alpha,\beta,x,y)$ 
& $O((\beta-\alpha+1)r)$
& $O(\beta-\alpha+\lg\sigma)$ \\ \hline

$\labrnk(\alpha,x,y)$ 
& $O(\alpha r)$
& $O(\alpha+\lg\sigma)$\\ \hline

$\labsel(\alpha,j,x,y)$ 
& $O((\sigma-\alpha+1)r)$
& $O(j\lg\sigma)$\\ \hline

$\labacc(\alpha,\beta,x,y)$ 
& $O((\beta-\alpha+1)r)$ 

  $O((y-x+1)a\lg\alpha)$
& $O((k+1)\lg\sigma)$\\ \hline

$\labmin(\alpha,x,y)$ 
& $O((\sigma-\alpha+1)r)$

  $O((y-x+1)a\lg\alpha)$
& $O(\lg\sigma)$\\ \hline

$\objnum(\alpha,\beta,x,y)$ 
& 
  $O((y-x+1)a\lg\alpha)$
& $O((y-x+1)\lg\sigma)$ \\ \hline 

$\objrnk(\alpha,\beta,x)$ 
& 
  $O(xa\lg\alpha)$
& $O(x\lg\sigma)$  \\ \hline 

$\objsel(\alpha,\beta,x,j)$ 
&  $O((n-x+1)a\lg\alpha)$
& $O(j\lg\sigma)$\\ \hline

$\objacc(\alpha,\beta,x,y)$ 
& $O((\beta-\alpha+1)(r+sk))$

  $O((y-x+1)a\lg\alpha)$
& $O((k+1)\lg\sigma)$\\ \hline

$\objmin(\alpha,\beta,x)$ 
& $O((\beta-\alpha+1)(r+s))$

  $O((n-x+1)a\lg\alpha)$
& $O(\lg\sigma)$\\ \hline

$\labrnko(\alpha,x)$ 
& 

  $O(a\lg\alpha)$
& $O(\lg\sigma)$\\ \hline

$\labselo(\alpha,j,x)$ 
& 

  $O(a \lg\alpha)$
& $O(\lg\sigma)$\\ \hline

$\labmino(\alpha,x)$ 
& 

  $O(a \lg\alpha)$
& $O(\lg\sigma)$\\ \hline

$\labacco(\alpha,\beta,x)$ 
& 

  $O(a (k+\lg\alpha))$
& $O((k+1)\lg\sigma)$\\ \hline

$\objrnko(\alpha,x)$ 
& $O(r)$ 
& $O(\lg\sigma)$\\ \hline

$\objselo(\alpha,x,j)$ 
& $O(r+s)$  
& $O(\lg\sigma)$ \\ \hline

$\objmino(\alpha,x)$ 
& $O(r+s)$  
& $O(\lg\sigma)$ \\ \hline

$\objacco(\alpha,x,y)$ 
& $O(r+sk)$  
& $O((k+1)\lg\sigma)$ \\ \hline
\end{tabular}
}\caption{Time complexity for the operations using
{\sc BinRel-Str} and {\sc BinRel-WT}. The parameter $k$ represents the size
of the output for the $\access$ operators; one can consider $k=1$ for the reductions given in
Theorem \ref{thm:reductions}.}
\label{tab:time_wt_vs_str}
\end{center}
\end{table}

Various string representations \cite{GMR06,BGNN10} offer times $a$, $r$,
and $s$ that are constant or log-logarithmic on $\sigma$. These yield the
best time complexities we know of for the row-wise and column-wise operations,
although these form a rather limited subset of the operations we have defined.

The space used by techniques based on representing $B$ and $S$ (including
{\sc BinRel-WT} and {\sc BinRel-GWT}) is usually unrelated to $H(\R)$, the
entropy of the binary relation. Various representations for $S[1,t]$ achieve 
space $tH_0(S)$ plus some redundancy \cite{GGV03,GMR06,BGNN10}. This is 
$tH_0(S) = \sum_{\alpha \in [1,\sigma]} n_\alpha \lg \frac{t}{n_\alpha}$,
where $n_\alpha$ is the number of pairs of the form $(\alpha,\cdot)$ in $\R$.
While this can be lower than $H(\R)$ (which shows that our measure
$H(\R)$ is rather crude), it can also be arbitrarily higher. For example an
almost full binary relation has an entropy $H(\R)$ close to zero, but its
$tH_0(S)$ is close to $n\sigma\lg\sigma$. A clearer picture is obtained if we
assume that $S$ is represented in plain form using $t\lg\sigma$ bits. This
is to be compared to $H(\R) = t\lg\frac{n\sigma}{t} + O(t)$, which shows that
the string representation is competitive for sparse relations, $t=O(n)$.

\section{Using Wavelet Trees: {\sc BinRel-WT}}
\label{sec:binrelwt}

Among the many string representations of $S$ we can choose in {\sc BinRel-Str}
scheme, wavelet trees \cite{GGV03} turn out to be particularly interesting.
Although the time wavelet trees offer for $a$, $r$ and $s$ is $O(\lg\sigma)$,
not the best 
ones for large $\sigma$, wavelet trees allow one to support many more operations
efficiently, via other algorithms than those used by the three basic operations.
We call this representation {\sc BinRel-WT}. Table \ref{tab:time_wt_vs_str} 
summarizes the time complexity for each operation using {\sc BinRel-WT}, in
comparison to a general {\sc BinRel-Str}.  Next, we show how to support some 
key operations efficiently; the other complexities are inferred from 
Theorem \ref{thm:reductions}.

The first lemma states a well-known algorithm on wavelet trees
\cite{MNtcs07}.

\begin{lemma}
\label{lem:wtop1}
{\sc BinRel-WT} supports $\relrnk(\alpha,x)$ in $O(\lg\sigma)$ time.
\end{lemma}

\begin{proof}
This is $\rank_{\le \alpha}(S,\map(x))$, where operation
$\rank_{\le\alpha}(S,p)$ counts the number of symbols $\le\alpha$ in
$S[1,p]$.  It can be supported in time $O(\lg\sigma)$ on the wavelet tree of
$S$ by following the root-to-leaf branch corresponding to $\alpha$, while
counting at each node the number of objects preceding position $p$ that are
related with a label preceding $\alpha$, as follows.  Start at the root $v$
with counter $c \leftarrow 0$.  If $\alpha$ corresponds to the left
subtree, then enter the left subtree with $p \leftarrow rank_0(B_v,p)$.
Else enter the right subtree with $c \leftarrow c + rank_0(B_v,p)$ and $p
\leftarrow \rank_1(B_v,p)$. Continue recursively until a leaf is reached 
(indeed, that of $\alpha$), where the answer is $c+p$.
\end{proof}

The next lemma solves an extended variant of a query called
{\em range\_quantile} in the literature, which was also solved with wavelet
trees within the same complexity \cite{GNP11}. Note that the lemma gives
also a solution within the same time complexity for $\labmin$, which in
the literature \cite{GNP11} was called {\em range\_next\_value} and
solved with an ad-hoc algorithm, within the same time.

\begin{lemma}
\label{lem:wtop2}
{\sc BinRel-WT} supports $\relsellabfst(\alpha,j,x,y)$ in $O(\lg\sigma)$
time.
\end{lemma}

\begin{proof}
We first get rid of $\alpha$ by setting $j \leftarrow
j+\relnum(1,\alpha-1,x,y)$ and thus reduce to the case $\alpha = 1$.
Furthermore we map $x$ and $y$ to the domain of $S$ by $p \leftarrow
\map(x-1)+1$ and $q \leftarrow \map(y)$.  We first find which is the symbol
$\beta$ whose row contains the $j$-th element.  For this sake we first find
the $\beta$ such that $\rank_{\le\beta-1}(S,p,q) < j \le
\rank_{\le\beta}(S,p,q)$. This is achieved in time $O(\lg\sigma)$ as
follows.  Start at the root $v$ and set $j' \leftarrow j$.  If
$\rank_0(B_v,p,q) \ge j$, then continue to the left subtree with $p
\leftarrow \rank_0(B_v,p-1)+1$ and $q \leftarrow \rank_0(B_v,q)$.  Else
continue to the right subtree with $j' \leftarrow j' - \rank_0(B_v,p,q)$,
$p \leftarrow \rank_1(B_v,p-1)+1$, and $y \leftarrow \rank_1(B_v,q)$.  The
leaf arrived at is $\beta$.  Finally, we 
answer $(\beta,\unmap(\select_\beta(S,j'+\rank_\beta(S,p-1))))$.
\end{proof}

The wavelet tree is asymmetric with respect to objects and labels. The
transposed problem, $\relselobjfst$, turns out to be harder. We present,
however, a polylogarithmic-time solution.

\begin{lemma}
\label{lem:wtop3}
{\sc BinRel-WT} supports $\relselobjfst(\alpha,\beta,x,j)$  in $O(\min(\lg
n,\lg j\lg(\beta-\alpha+1))\lg\sigma)$ time.
\end{lemma}

\begin{proof}
Remember that the elements are written in $S$ in object major order.
First, we note that the particular case where $[\alpha,\beta] = [1,\sigma]$
is easily solved in $O(\lg\sigma)$ time, by doing $j' \leftarrow
j+\relnum(1,\sigma,1,x-1)$ and returning $(S[j'],\unmap(j'))$.  In the
general case, one can obtain time $O(\lg n \lg\sigma)$ by binary
searching the column $y$ such that $\relnum(\alpha,\beta,x,y-1) < j \le
\relnum(\alpha,\beta,x,y)$.  Then the answer is 
$(\labselo(\alpha,j-\relnum(\alpha,\beta,x,y-1),y),y)$ (note that
Lemma~\ref{lem:wtop2} already gives us $\labselo$ in time $O(\lg\sigma)$).  

To obtain the other complexity, we find the $O(\lg(\beta-\alpha+1))$ wavelet 
tree nodes that cover the interval $[\alpha,\beta]$; let these be $v_1,~v_2,
\ldots, v_k$. We map position $p=\map(x-1)+1$ from the root towards those 
$v_i$s, obtaining all the mapped positions $p_i$ in $O(k+\lg\sigma)$ time.
\cite{GNP11} 
Now the answer is within the positions $[p_i,p_i+j-1]$ of some $i$. We
cyclically take each $v_i$, choose the middle element of its interval, and
map it towards the root, obtaining position $q$, corresponding to pair
$(S[q],\unmap(q))$. If $\relrnkobjfst(\alpha,\beta,S[q],\unmap(q)) -
\relnum(\alpha,\beta,1,x-1) = j$, the answer is $(S[q],\unmap(q))$.
Otherwise we know whether $q$ is before or after the answer. So we discard
the left or right interval in $v_i$. After $O(k \lg j)$ such iterations we
have reduced all the intervals of length $j$ of all the nodes $v_i$,
finding the answer. Each iteration costs $O(\lg\sigma)$ time.
\end{proof}

The next lemma solves a more general variant of a problem that was called
{\em prevLess} in the literature, and also solved with wavelet trees 
\cite{KN11}. We achieve the same complexity for this more general variant.
Note this is a simplification of $\relselobjfst$ that we can solve within
time $O(\lg\sigma)$, whereas for general $j$ we cannot.

\begin{lemma}
\label{lem:wtop4}
{\sc BinRel-WT} supports $\relminobjfst(\alpha,\beta,\gamma,x)$ in 
$O(\lg\sigma)$ time per pair output.
\end{lemma}

\begin{proof}
We first use $\labmino(\gamma,x)$ (which we already can solve in time
$O(\lg\sigma)$ as a consequence of Lemma~\ref{lem:wtop2}) to search for
a point in the band $[\gamma,\beta] \times [x,x]$. If we find one, then this
is the answer, otherwise we continue with the area $[\alpha,\beta] \times 
[x+1,n]$.

Just as for the second solution of Lemma~\ref{lem:wtop3}, we obtain the
positions $p_i$ at the nodes $v_i$ that cover $[\alpha,\beta]$.  The first
element to deliver is precisely one of those $p_i$.  We have to merge the
results, choosing always the smaller, as we return from the recursion that
identifies the $v_i$ nodes.  If we are in $v_i$, we return $q = p_i$.
Else, if the left child of $v$ returned $q$, we map it to $q' \leftarrow
\select_0(B_v,q)$.  Similarly, if the right child of $v$ returned $q$, we map
it to $q'' \leftarrow \select_1(B_v,q)$.  If we have only $q'$ ($q''$), we
return $q=q'$ ($q=q''$); if we have both we return $q=\min(q',q'')$. The
process takes $O(\lg\sigma)$ time.  When we arrive at the root we have the
next position $q$ where a label in $[\alpha,\beta]$ occurs in $S$, and thus
return $\unmap(q)$.
\end{proof}

The next result can also be obtained by considering the complexity of the
{\sc BinRel-Str} scheme implemented over a wavelet tree.

\begin{lemma}
\label{lem:wtop5}
{\sc BinRel-WT} supports $\objselo(\alpha,x,j)$ in $O(\lg\sigma)$ time.
\end{lemma}

\begin{proof}
This is a matter of selecting the $j$-th occurence of the label $\alpha$ in
$S$, after the position of the pair $(\alpha,x)$. The formula is
$\unmap(\select_\alpha(S,j+\objrnko(\alpha,x-1)))$.
\end{proof}

The next operation is the first of the set we cannot solve within
polylogarithmic time.

\begin{lemma}
\label{lem:wtop6}
{\sc BinRel-WT} supports $\labnum(\alpha,\beta,x,y)$ in
$O(\beta-\alpha+\lg\sigma)$ time.
\end{lemma}

\begin{proof}
After mapping $[x,y]$ to positions $S[p,q]$, we descend in the wavelet
tree to find all the leaves in $[\alpha,\beta]$ while remapping $[p,q]$
appropriately. We count one more label each time we arrive at a leaf, and
we stop descending from an internal node if its range $[p,q]$ is empty.
The complexity comes from the number of wavelet tree nodes accessed to
reach such leaves \cite{GNP11}.
\end{proof}

\begin{figure}[t]
\begin{center}
\includegraphics[width=11cm]{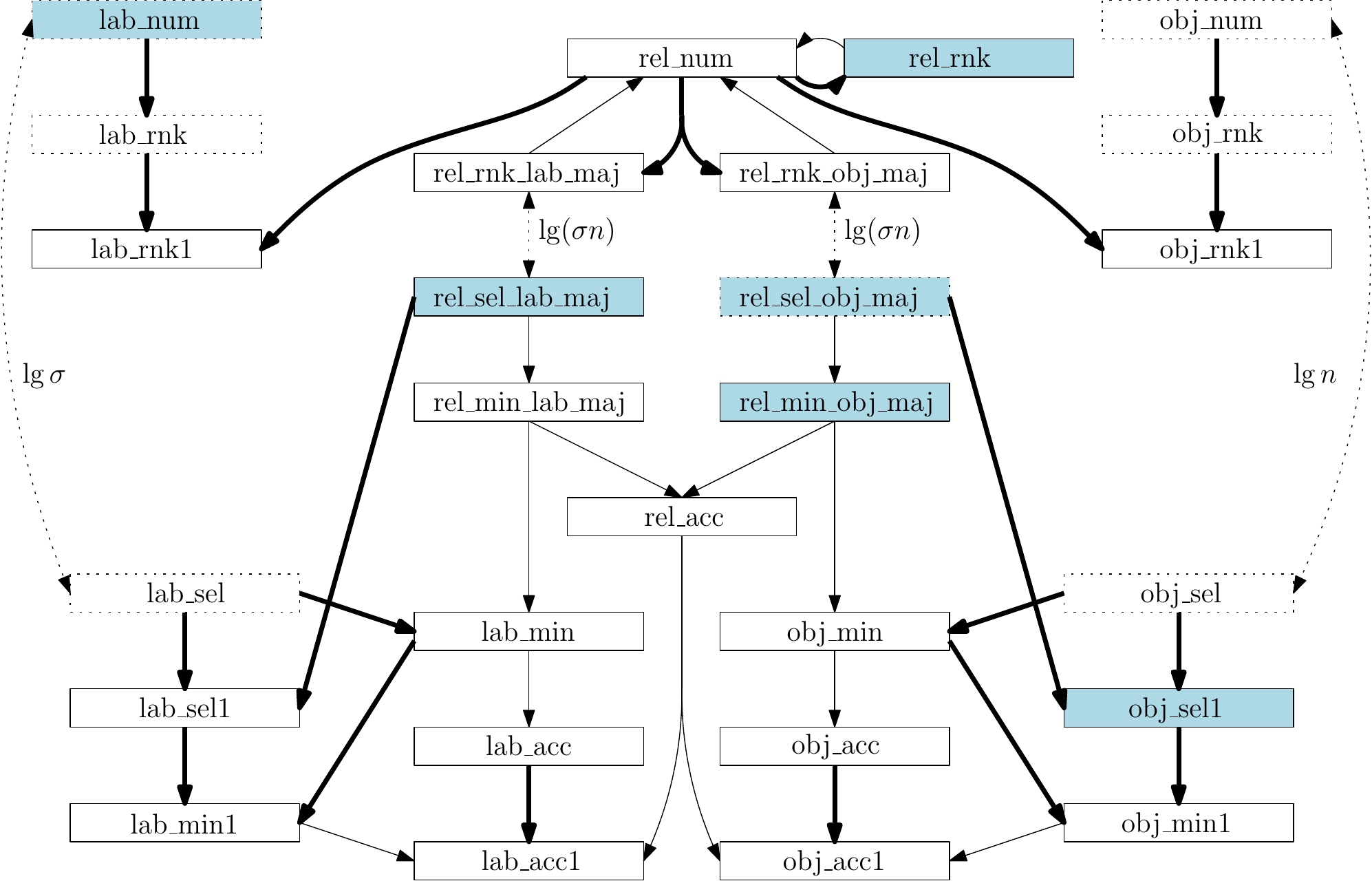}
\caption{Reductions among operations supported by {\sc BinRel-WT}. The
dotted boxes are operations supported in $\omega(\lg\sigma)$ time, the
filled boxes represent operations directly addressed in the paper, and the
blank boxes are supported via reductions.}
\label{fig:reductions_wt}
\end{center}
\end{figure}

The remaining operations are solved naively: $\labsel$ and $\objsel$ use,
respectively, $\labmin$ and $\objmin$ successively, and $\objnum$ and 
$\objrnk$ use $\objrnko$ successively,

The overall result is stated in the next theorem and illustrated in
Figure~\ref{fig:reductions_wt}.

\begin{theorem}
The structure {\sc BinRel-WT}, for a binary relation $\R$ of $t$ pairs over
$[1,\sigma] \times [1,n]$, uses $t\lg\sigma+O(n+t)$ bits of space and
supports the operations within the time complexities given in 
Table~\ref{tab:time_wt_vs_str}. 
\label{thm:binrelwt}
\end{theorem}

\begin{proof} 
The space assumes a plain uncompressed wavelet tree and bitmap representations,
and the time complexities have been obtained throughout the section.
\end{proof}

\section{Using a Generalized Wavelet Tree: {\sc BinRel-GWT}}

The results we obtained for the wavelet tree can be extended to the generalized
wavelet tree, improving complexities in many cases (recall 
Section~\ref{sec:seqlarge}). We refer to the structure 
that represents $S$ using the generalized wavelet tree as {\sc BinRel-GWT},
and we use $\mu$ to represent the fan-out of the tree. We require
$\mu \in O(\lg^\epsilon (n\sigma))$, assuming that the RAM machine can 
address up to $n \times \sigma$ cells. To simplify we will assume 
$\sigma \le n$ and then use simply $\mu \in O(\lg^\epsilon n)$.

A first simple result stems from the fact that the string operations are sped
up on generalized wavelet trees.

\begin{lemma}
\label{lem:gwtop1}
{\sc BinRel-GWT} supports $\objrnko$ and $\objselo$ in time $O(\lg_\mu
\sigma)$ for any $\mu\in\Theta(\lg^{\epsilon} n)$ and any constant $0<\epsilon<1$.
\end{lemma}

\begin{proof}
This follows directly from the results on general {\sc BinRel-Str} structures.
\end{proof}

The following notation will be useful to describe our algorithms within
wavelet tree nodes. Note that all are easily computed in constant time. 

\begin{itemize}  
\item $\child(k)$: Given a symbol $k\in[1,\mu]$, this is the subtree labeled
$k$ of the current node.
\item $\g(\alpha)$: Given a symbol $\alpha\in[1,\sigma]$, $\g(\alpha)$ is the 
symbol $k$ such that $\child(k)$ contains the leaf corresponding to $\alpha$. 
\item $\g^{-1}(k)$: Given a symbol $k\in[1,\mu]$, $\g^{-1}(k) =
  \min\{\alpha, k=\g(\alpha)\}$. 
\end{itemize}

The next lemma shows how to speed up all the range counting operations.
The result is known in the literature for $n\times n$ grids, even within
$n\lg n (1+o(1))$ bits of space \cite{BHMM09}.

\begin{lemma}
\label{lem:gwtop1.5}
{\sc BinRel-GWT} supports $\relrnk(\alpha,x)$ in $O(\lg_\mu \sigma)$
time, for any $\mu\in\Theta(\lg^{\epsilon} n)$ and any constant $0<\epsilon<1$.
\end{lemma}

\begin{proof}
As in the case of {\sc BinRel-WT}, we reduce this problem to the one of
computing $\rank_{\leq\alpha}(S,\map(x))$, which can be done by following a
similar procedure: We follow the path for $\alpha$ starting at the root in
position $p=\map(x)$ and with a counter $c \leftarrow 0$. Every time we move 
to a subtree, we increase $c \leftarrow c + \rank_{\leq g(\alpha)-1}(S_v,p)$.
When we arrive at the leaf, the answer is $c+p$.

Operation $\rank_{\leq k}(S_v,p)$ can be solved in constant time for
$\mu\in\Theta(\lg^{\epsilon} n)$ analogously as done for $\rank_k$ on small
alphabets (recall Section~\ref{sec:seqsmall}). We store for each 
$k \in [1,\mu]$ a bitmap $B_{\le k}$ such that $B_{\le k}[i] = 1$ iff 
$S_v[i] \le k$. Thus $\rank_{\le k}(S_v,p) = \rank_1(B_{\le k},p)$ is computed 
in constant time and the whole process takes $O(\lg_\mu \sigma)$ time.
\end{proof}

The next lemma covers operation $\relsellabfst$, on which we cannot improve
upon the complexity given by {\sc BinRel-WT}. Note this means that
$O(\lg\sigma)$ is still the best time complexity for supporting {\em 
range\_quantile} queries within linear space \cite{GNP11}.

\begin{lemma}
\label{lem:gwtop2}
{\sc BinRel-GWT} supports $\relsellabfst(\alpha,j,x,y)$ in
$O(\lg\sigma)$ time.
\end{lemma}

\begin{proof}
This is solved in a similar way to the one presented for {\sc
BinRel-WT}. We find $v$ such that $\rank_{\leq \beta-1}(S,p,q) < v
\leq \rank_{\leq \beta}(S,p,q)$. The only difference is that in this case
we have to do, at each node, a binary search for the right child $k\in[1,\mu]$ 
to descend, and thus the time is $O(\lg \mu \lg_\mu \sigma ) = O(\lg \sigma)$.
\end{proof}

For the next operations we will augment the generalized wavelet tree with a 
set of bitmaps inside each node $v$. More specifically, we will add
$\mu(\mu+1)/2$ bitmaps $B_{k,l}$, where $B_{k,l}[i]=1$ iff $S_v[i] \in [k,l]$.
Just as bitmaps $B_{\le k}$, bitmaps $B_{k,l}$ are not represented explicitly,
but only their index is stored (recall Section~\ref{sec:seqsmall}), and their
content is simulated in constant time using $S_v$. Their total space for
a sequence $S_v[1,n]$ is $O(n \mu^2 \lg\lg n / \lg_\mu n)$. To make this
space negligible, that is, $o(n\lg\mu)$, it is sufficient that 
$\mu = O(\lg^\epsilon n)$ for any constant $0<\epsilon < 1/2$. (A related idea
has been used by Farzan et al.~\cite{FGN10}.)

The next lemma shows that the current solution for operation {\em prevLess} 
\cite{KN11} can be sped up by an $O(\lg\lg n)$ factor.

\begin{lemma} \label{lem:gwtop3} 
{\sc BinRel-GWT} supports $\relminobjfst(\alpha,\beta,\gamma,x)$ in 
$O(\lg_\mu \sigma)$ time, for any $\mu\in\Theta(\lg^{\epsilon} n)$ and any
constant $0<\epsilon<1/2$. 
\end{lemma}

\begin{proof}
We first run query $\relminobjfst(\gamma,\beta,\gamma,x)$, and if the result
is on column $x$, we report it. Otherwise we run query 
$\relminobjfst(\alpha,\beta,\alpha,x+1)$. This means that we can focus on
a simpler query of the form $\relminobjfst(\alpha,\beta,x)$, which finds the
first pair in $[\alpha,\beta] \times [x,n]$, in object-major order.
We map $[x,n]$ to $S[p,t]$ as usual and then proceed recursively on the wavelet
tree, remapping $p$. At each node $v$, we decompose the query into three 
subqueries, and then take the minimum result of the three:

\begin{enumerate}
\item $\relminobjfst(\alpha, \g^{-1}(\g(\alpha)+1)-1, x)$ on node 
$\child(\alpha)$;
\item $\relminobjfst(\g^{-1}(\g(\alpha)+1),\g^{-1}(\g(\beta))-1, x)$ on the
same node $v$; 
\item $\relminobjfst(\g^{-1}(\g(\beta)),\beta, x)$ on node $\child(\beta)$.
\end{enumerate}

Note that queries of type 1 will generate, recursively, only $O(\lg_\mu \sigma)$
further queries of type 1 and 2, and similarly queries of type 3 will generate
$O(\lg_\mu \sigma)$ further queries of type 3 and 2. The only queries that 
actually deliver values are those of type 2, and we will have to take the 
minimum over $O(\lg_\mu \sigma)$ such results.

A query of type 2 is solved in constant time using bitmap 
$B_{\g(\alpha)+1, \g(\beta)-1}$, by computing $q = 
\select_1(B_{\g(\alpha)+1,\g(\beta)-1},\rank_1(B_{\g(\alpha)+1,\g(\beta)-1},p-1)+1)$. 
This returns a position $S_v[q]$. As we return from the recursion, we remap
$q$ in its parent in the usual way, and then (possibly) compare $q$ with the 
result of a query of type 1 or 3 carried out on the parent. We keep the
minimum $q$ value along the way, and when we arrive at the root we return
$(S[q],\unmap(q))$.
\end{proof}

For the next lemma we need a further data structure. For each sequence
$S_v[1,n]$, we store an RMQ structure (Section~\ref{sec:rmq}), using $O(n) =
o(n\lg\mu)$ bits and finding in constant time position of a minimum symbol in
any range $S_v[i,j]$. This results improves upon the result for query
{\em range\_next\_value} \cite{GNP11}.

\begin{lemma} \label{lem:gwtop3.5} 
{\sc BinRel-GWT} supports $\relminlabfst(\alpha,x,y,z)$ in 
$O(\lg_\mu \sigma + \lg \mu)$ time, for any $\mu\in\Theta(\lg^{\epsilon} n)$ and 
any constant $0<\epsilon<1$. 
\end{lemma}

\begin{proof}
Again, we can focus on a simpler query $\relminlabfst(\alpha,x,y)$. We map
$[x,y]$ to $S[p,q]$ as usual, and the goal is to find the leftmost minimum 
symbol in $S[p,q]$ that is larger than $\alpha$. 

Assume we are in a wavelet tree node $v$ and the current interval of interest
is $S_v[p,q]$. Then, if $S_v[p,q]$ contains symbol $\g(\alpha)$ (which is known
in constant time with $\rank_{\g(\alpha)}(S_v,p,q) > 0$), we have to
consider it first, by querying recursively the child labeled $\g(\alpha)$.
If this recursive call returns an answer $p'$, we return it in turn, remapping 
it to the parent node. If it does not, then any symbol larger than $\alpha$ in
the range must correspond to a symbol strictly larger than $\g(\alpha)$ in
$S_v[p,q]$. We check in constant time whether there is any value larger than 
$\g(\alpha)$ in $S_v[p,q]$, using $\rank_{\le \g(\alpha)}(S_v,p,q) < q-p+1$.
If there is none, we return in turn with no answer.

If there is an answer, we binary search for the smallest 
$k \in [\g(\alpha)+1,\mu]$ such that $\rank_{\le k}(S_v,p,q)  >
\rank_{\le \g(\alpha)}(S_v,p,q)$. This binary search takes $O(\lg\mu)$ time
and is done only once along the whole process. Once we identify the right $k$,
we descend to the appropriate child and start the final stage of the process.

The final stage starts at a node where all the local symbols represent
original symbols that are larger than $\alpha$, and therefore we simply look
for the position $m=\textsc{rmq} (S_v,p,q)$, which gives us, in constant time,
the first occurrence of the minimum symbol in $S_v$, and descend to child
$S[m]$. This is done until reaching a leaf, from where we return to the root,
at position $p'$, and return $(S[p'],\unmap(p'))$. It is easy to see that we 
work $O(1)$ time on $O(\lg_\mu \sigma)$ nodes and $O(\lg \mu)$ once.
\end{proof}

\begin{lemma}
\label{lem:gwtop4}
{\sc BinRel-GWT} supports $\relselobjfst(\alpha,\beta,x,j)$ in 
$O(\min(\lg n,\lg j\lg(\beta-\alpha+1))\lg_\mu\sigma)$ time, for any
$\mu\in\Theta(\lg^{\epsilon} n)$ and any constant $0<\epsilon<1/2$.
\end{lemma}

\begin{proof}
The complexities are obtained the same way as for {\sc BinRel-WT}. The binary 
search over $\relnum$ is sped up because {\sc BinRel-GWT} supports this 
operation faster. The other complexity is in principle higher, because the
interval $[\alpha,\beta]$ is split into as much as $O(\mu\lg(\beta-\alpha+1))$
nodes. However, this can be brought down again to $O(\lg(\beta-\alpha+1))$ by
using the parent node $v$ of each group of (up to $\mu$) contiguous leaves
$[k,l]$, and using $\select_1$ on the bitmaps $B_{k,l}$ of those parent nodes
in order to simulate a contiguous range with all the values in $[k,l]$. So
we still have $O(\lg(\beta-\alpha+1))$ binary searches of $O(\lg j)$ steps,
and now each step costs $O(\lg_\mu\sigma)$.
\end{proof}

\begin{lemma}
\label{lem:gwtop5}
{\sc BinRel-GWT} supports $\labnum(\alpha,\beta,x,y)$ in
$O(\beta-\alpha+\lg_{\mu}\sigma)$ time, for any $\mu\in\Theta(\lg^{\epsilon} n)$ and any constant
$0<\epsilon<1/2$.
\end{lemma}

\begin{proof}
We follow the same procedure as for {\sc BinRel-WT}. The main difference is 
how to compute the nodes covering the range $[\alpha,\beta]$. This can be done 
in a na\"ive way by just verifying whether each symbol appears in the range
of $S_v$, but this raises the complexity by a factor of $\mu$. Thus we need a
method to list the symbols appearing in a range of $S_v$ without probing 
non-existent ones. We resort to a technique loosely inspired by Muthukrishnan
\cite{Mut02}. To list the symbol from a range $[k,l]$ that exist in
$S_v[p,q]$, we start with the first symbol of the range that appears in
$S_v[p,q]$. This is obtained with 
$p'=\select_1(B_{k,l},\rank_1(B_{k,l},p-1)+1)$. If $p'>q$ then there are no
such symbols. Otherwise, let $k' = S_v[p']$.
Then we know that $k'$ appears in $S_v[p,q]$. Now we continue recursively
with subranges $[k,k'-1]$ and $[k'+1,l]$. The recursion stops when no $p'$ is
found, and it yields all the symbols appearing in $S_v[p,q]$ in $O(1)$ time
per symbol.
\end{proof}

\begin{figure}[t]
\begin{center}
\includegraphics[width=11cm]{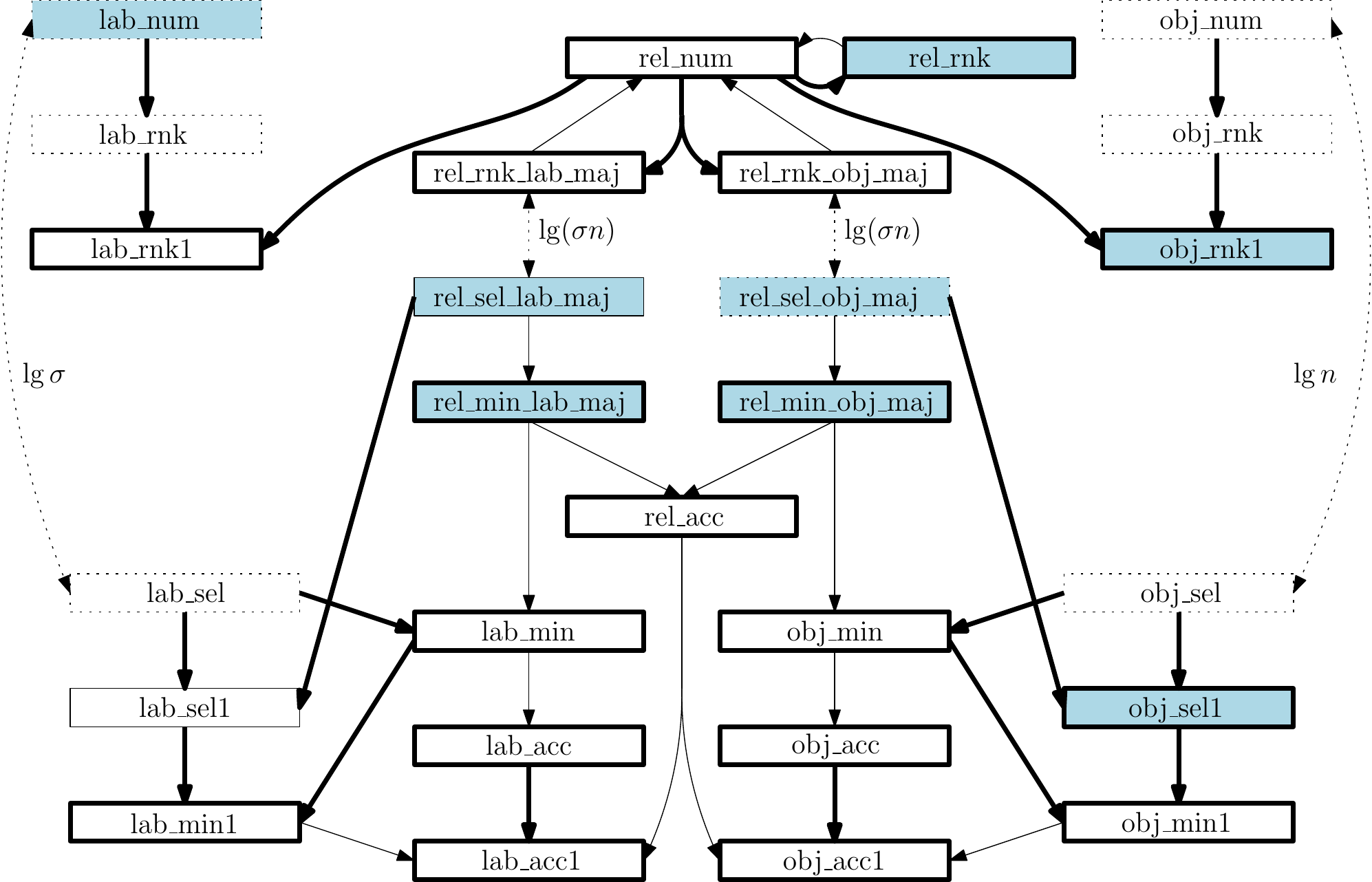}
\caption{Reductions among operations supported by {\sc BinRel-GWT}. The
dotted boxes are operations supported in $\omega(\lg\sigma)$ time, the
solid ones in $O(\lg\sigma)$ time, and the thick ones in time
$O(\lg_\mu\sigma)$ or $O(\lg_\mu \sigma + \lg\mu)$. The
filled boxes represent operations addressed directly, and the
blank boxes are supported via reductions.}
\label{fig:reductions_gwt}
\end{center}
\end{figure}

The remaining operations are obtained by brute force, just as with {\sc
BinRel-WT}. Figure~\ref{fig:reductions_gwt} illustrates the reductions used.

\begin{theorem}
The structure {\sc BinRel-GWT}, for a binary relation $\R$ of $t$ pairs over 
$[1,\sigma] \times [1,n]$, requires $t\lg\sigma (1+o(1)) + O(n+t)$ bits of 
space and supports the operations within the time complexities given in
Table \ref{tab:wt_vs_gwt}.
\label{thm:binrelgwt}
\end{theorem}

\begin{table}[tb]
\begin{center}
{\scriptsize 
\begin{tabular}{| l | p{3.9cm} | p{3.3cm} |} \hline
Operation & {\sc BinRel-GWT} & {\sc BinRel-WT}\\
\hline\hline
%
%
$\relnum(\alpha,\beta,x,y)$ 
& $O(\lg\sigma/\lg\lg n)$
& $O(\lg\sigma)$ \\ \hline

$\relrnk(\alpha,x)$ 
& $O(\lg\sigma/\lg\lg n)$
& $O(\lg\sigma)$ \\ \hline

$\relrnklabfst(\alpha,x,y,z)$ 
& $O(\lg\sigma/\lg\lg n)$
& $O(\lg\sigma)$\\ \hline

$\relsellabfst(\alpha,j,x,y)$ 
& $O(\lg\sigma)$
& $O(\lg\sigma)$\\ \hline

$\relminlabfst(\alpha,x,y,z)$ 
& $O(\lg\sigma/\lg\lg n + \lg\lg n)$
& $O(\lg\sigma)$\\ \hline

$\relrnkobjfst(\alpha,\beta,\gamma,x)$ 
& $O(\lg\sigma/\lg\lg n)$
& $O(\lg\sigma)$\\ \hline

$\relselobjfst(\alpha,\beta,x,j)$ 
& $O(\lg j\lg(\beta-\alpha+1)\lg\sigma/\lg\lg n)$

  $O(\lg n \lg\sigma/\lg \lg n)$
& $O(\lg j\lg(\beta-\alpha+1)\lg\sigma)$

  $O(\lg n \lg\sigma)$ \\ \hline

$\relminobjfst(\alpha,\beta,\gamma,x)$ 
& $O(\lg\sigma/\lg\lg n)$
& $O(\lg\sigma)$\\ \hline

$\relacc(\alpha,\beta,x,y)$ 
& $O((k+1)\lg\sigma/\lg\lg n)$
& $O((k+1)\lg\sigma)$\\ \hline

$\labnum(\alpha,\beta,x,y)$ 
& $O(\beta-\alpha+\lg\sigma/\lg\lg n)$
& $O(\beta-\alpha+\lg\sigma)$ \\ \hline

$\labrnk(\alpha,x,y)$ 
& $O(\alpha+\lg\sigma/\lg\lg n)$
& $O(\alpha+\lg\sigma)$\\ \hline

$\labsel(\alpha,j,x,y)$ 
& $O(j(\lg\sigma/\lg\lg n + \lg\lg n))$
& $O(j\lg\sigma)$\\ \hline

$\labacc(\alpha,x,y)$ 
& $O((k+1)(\lg\sigma/\lg\lg n + \lg\lg n))$
& $O((k+1)\lg\sigma)$\\ \hline

$\labmin(\alpha,x,y)$ 
& $O(\lg\sigma/\lg\lg n + \lg\lg n)$
& $O(\lg\sigma)$\\ \hline

$\objnum(\alpha,\beta,x,y)$ 
& $O((y-x+1)\lg\sigma/\lg\lg n)$
& $O((y-x+1)\lg\sigma)$\\ \hline

$\objrnk(\alpha,\beta,x)$ 
& $O((y-x+1)\lg\sigma/\lg\lg n)$
& $O((y-x+1)\lg\sigma)$\\ \hline

$\objsel(\alpha,\beta,x,j)$ 
& $O(j\lg\sigma/\lg\lg n)$
& $O(j\lg\sigma)$\\ \hline

$\objacc(\alpha,\beta,x)$ 
& $O((k+1)\lg\sigma/\lg\lg n)$
& $O((k+1)\lg\sigma)$\\ \hline

$\objmin(\alpha,\beta,x)$ 
& $O(\lg\sigma/\lg\lg n)$
& $O(\lg\sigma)$\\ \hline

$\labrnko(\alpha,x)$ 
& $O(\lg\sigma/\lg\lg n)$
& $O(\lg\sigma)$\\ \hline

$\labselo(\alpha,j,x)$ 
& $O(\lg\sigma)$
& $O(\lg\sigma)$\\ \hline

$\labmino(\alpha,j,x)$ 
& $O(\lg\sigma/\lg\lg n + \lg\lg n)$
& $O(\lg\sigma)$\\ \hline

$\labacco(\alpha,j,x)$ 
& $O((k+1)\lg\sigma/\lg\lg n)$
& $O((k+1)\lg\sigma)$\\ \hline

$\objrnko(\alpha,x)$ & $O(\lg\sigma/\lg\lg n)$ & $O(\lg\sigma)$\\ \hline
$\objselo(\alpha,x,j)$ & $O(\lg\sigma/\lg\lg n)$  & $O(\lg\sigma)$ \\ \hline
$\objmino(\alpha,x,j)$ & $O(\lg\sigma/\lg\lg n)$  & $O(\lg\sigma)$ \\ \hline
$\objacco(\alpha,x,j)$ & $O((k+1)\lg\sigma/\lg\lg n)$  & $O((k+1)\lg\sigma)$ \\ \hline
\end{tabular}
%
}
\caption{Time complexity for the operations for
{\sc BinRel-GWT} and {\sc BinRel-WT}. The parameter $k$ represents the size
of the output for the $\access$ operators; one can consider $k=1$ for the reductions shown in
Theorem \ref{thm:reductions}.}
\label{tab:wt_vs_gwt}
\end{center}
\end{table}

\section{Binary Relation Wavelet Trees (BRWT)}
\label{sec:brwt}

\begin{figure}[t]
\begin{center}
\includegraphics[width=11cm]{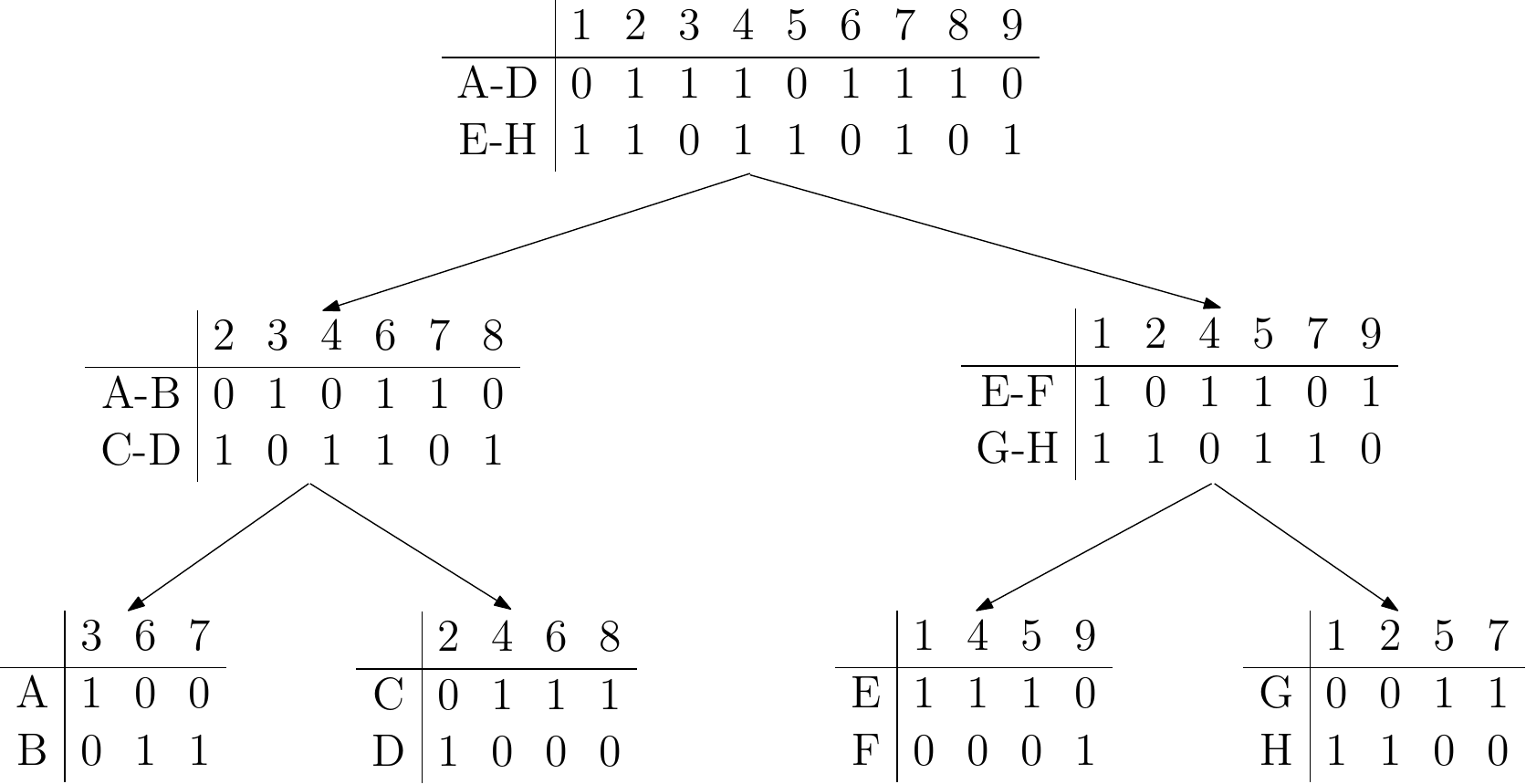}
\caption{Example of the BRWT for the binary relation in Figure \ref{fig:binrel_example}.}
\label{fig:br}
\end{center}
\end{figure}

We propose now a special wavelet tree structure tailored to the 
representation of binary relations. This wavelet tree contains two bitmaps 
per level at each node $v$, $B_v^l$ and $B_v^r$. At the root, $B_v^l[1,n]$ has 
the $x$-th bit set to $1$ iff there exists a pair $(\alpha,x)$ with $\alpha \in
[1,\lfloor\sigma/2\rfloor]$, and $B_v^r$ has the $x$-th bit set to $1$ iff
there exists a pair $(\alpha,x)$ with $\alpha \in
[\lfloor\sigma/2\rfloor+1,\sigma]$.  Left and right subtrees are
recursively built on the positions set to $1$ in $B_v^l$ and $B_v^r$,
respectively. The leaves (where no bitmap is stored) correspond to
individual rows of the relation.  We store a bitmap $B[1,\sigma+t]$
recording in unary the number of elements in each row. See
Figure~\ref{fig:br} for an example.  For ease of notation, we define
the following functions on $B$, trivially supported in constant-time:
$\lab(r) = 1+ \rank_0(B,\select_1(B,r))$ gives the label of the $r$-th pair
in a label-major traversal of $R$; while its inverse $\poslab(\alpha) =
\rank_1(B,\select_0(B,\alpha))$ gives the position in the traversal where
the pairs for label $\alpha$ start.

Note that, because an object $x$ may propagate both left and right, the
sizes of the second-level bitmaps may add up to more than $n$ bits.
Indeed, the last level contains $t$ bits and represents all the pairs
sorted in row-major order. As we will see, the BRWT has weaker functionality
than our former structures based on wavelet trees, but it reaches space 
proportional to $H(\R)$.

Lemmas \ref{lem:brwtop1} to \ref{lem:brwtop6} give a set of operations that
can be supported with the BRWT structure.

\begin{lemma}
\label{lem:brwtop1}
{\sc BRWT} supports $\relnum(\alpha,\beta,x,y)$ in
$O(\beta-\alpha+\lg\sigma)$ time.
\end{lemma}

\begin{proof}
We project the interval $[x,y]$ from the root to each leaf in
$[\alpha,\beta]$, adding up the resulting interval sizes at the leaves. Of
course we can stop earlier if the interval becomes empty. Note that we can
only count pairs at the leaves, not at internal nodes. 
\end{proof}

%

\begin{lemma}
\label{lem:brwtop3}
{\sc BRWT} supports $\relminlabfst(\alpha,x,y,z)$ in $O(\lg\sigma)$ time.
\end{lemma}

\begin{proof}
As before, we only need to consider the simpler query 
$\relminlabfst(\alpha,x,y)$. We reach the $O(\lg\sigma)$ wavelet tree nodes
$v_1, v_2, \ldots$ that cover the interval $[\alpha,\sigma]$, and map $[x,y]$ 
to all those nodes, in $O(\lg\sigma)$ time \cite{GNP11}. We choose the first 
such node, $v_k$, left to right, with a nonempty interval $[x,y]$. Now we 
find the leftmost leaf of $v_k$ that has a nonempty interval $[x,y]$, which is
easily done in $O(\lg\sigma)$ time. Once we arrive at such a leaf $\gamma$ with
interval $[x,y]$, we map $x$ back to the root, obtaining $x'$, and the answer
is $(\gamma,x')$.
\end{proof}

\begin{lemma}
\label{lem:brwtop4}
{\sc BRWT} supports $\relminobjfst(\alpha,\beta,\gamma,x)$ in $O(\lg\sigma)$ 
time.
\end{lemma}

\begin{proof}
As before, we only need to consider the simpler query 
$\relminobjfst(\alpha,\beta,x)$. 
Analogously to the proof of Lemma~\ref{lem:wtop4}, we cover $[\alpha,\beta]$ 
with $O(\lg\sigma)$ wavelet tree nodes $v_1, v_2, \dots$, and map $x$ to $x_i$
at each such $v_i$, all in $O(\lg\sigma)$ time.  Now, on the way back of
this recursion, we obtain the smallest $y \ge x$ in the root associated to some
label in $[\alpha,\beta]$. In this process we keep track of the node $v_i$
that is the source of $y$, preferring the left child in case of ties. Finally, 
if we arrive at the root with a value $y$ that came from node $v_i$, we start 
from position $x'=x_i$ at node $v_i$ and find the leftmost leaf of $v_i$ 
related to $y$. This is done by going left whenever possible (i.e., if 
$B_v^l[x']=1$) and right otherwise, and remapping $x'$ appropriately at each 
step. Upon reaching a leaf $\gamma$, we report $(\gamma,y)$. 
\end{proof}

\begin{lemma}
\label{lem:brwtop5}
{\sc BRWT} supports $\objselo(\alpha,x,j)$ in $O(\lg\sigma)$ time.
\end{lemma}

\begin{proof}
We map $x-1$ from the root to $x'$ in leaf $\alpha$, then walk upwards the
path from $x'+j$ to the root and report the position obtained.
\end{proof}

\begin{lemma}
\label{lem:brwtop6}
{\sc BRWT} supports $\labnum(\alpha,\beta,x,y)$ in
$O(\beta-\alpha+\lg\sigma)$ time.
\end{lemma}

\begin{proof}
We map $[x,y]$ from the root to each leaf in $[\alpha,\beta]$, adding one per
leaf where the interval is non-empty. Recursion can also stop when $[x,y]$
becomes empty.
\end{proof}

\begin{figure}[t]
\begin{center}
\includegraphics[width=11cm]{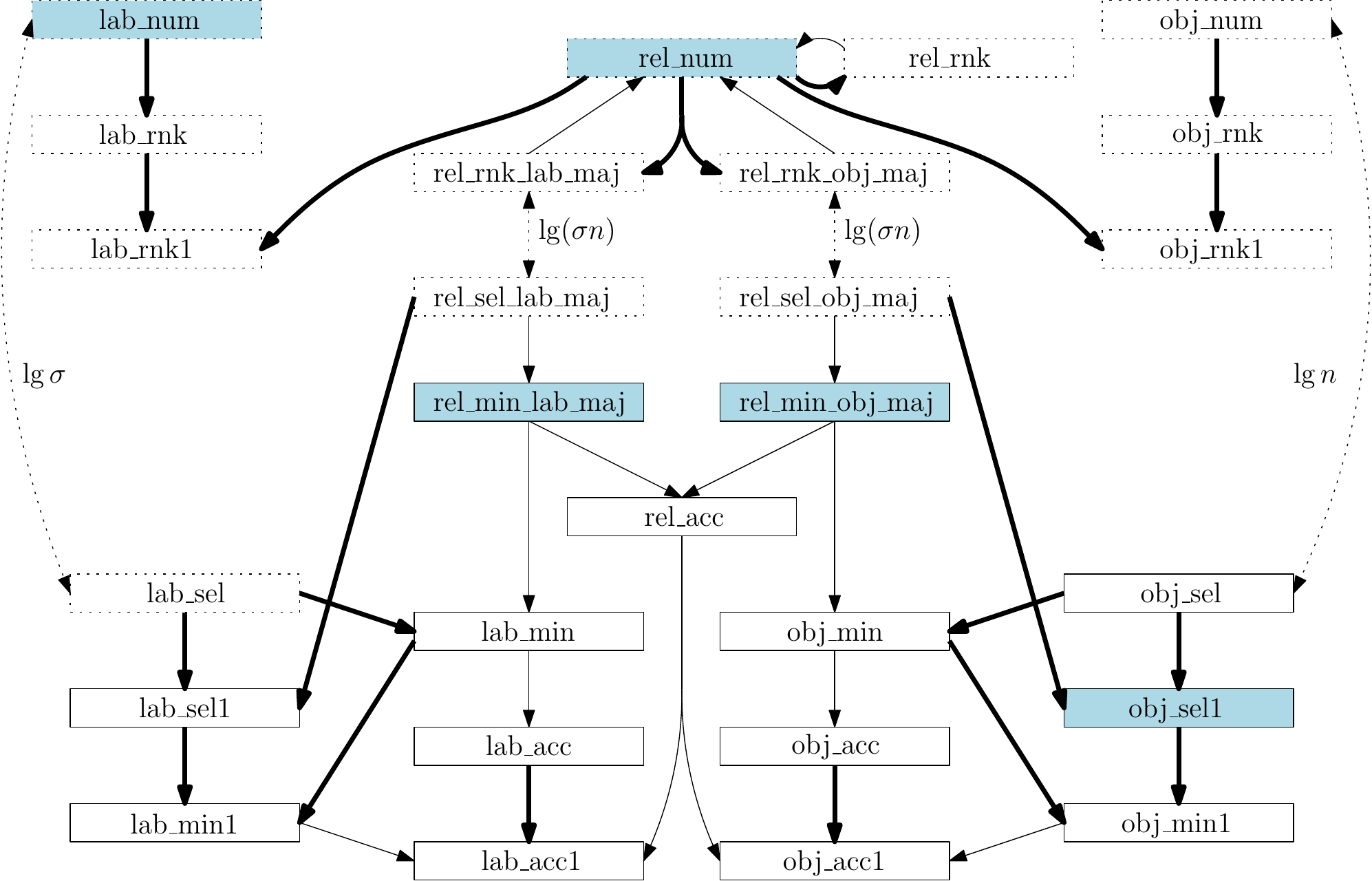}
\caption{Reductions among operations supported by BRWT. The dotted
boxes are operations supported in $\omega(\lg\sigma)$ time, the filled
boxes represent operations directly addressed in the paper, and the blank
boxes are supported via reductions.}
\label{fig:reductions_brwt}
\end{center}
\end{figure}

The remaining complexities are obtained by brute force: $\relsellabfst$ and
$\relselobjfst$ are obtained by iterating with $\relminlabfst$ and
$\relminobjfst$, respectively; and similarly $\labsel$, $\objsel$ and
$\labselo$ using $\labmin$, $\objmin$, and $\labmino$. Finally, as before
$\objnum$ and $\objrnk$ are obtained by iterating over $\objrnko$.
We have obtained the following theorem, illustrated in
Figure~\ref{fig:reductions_brwt}. 

\begin{theorem}
The {\sc BRWT} structure, for a binary relation $\R$ of $t$ pairs
over $[1,\sigma] \times [1,n]$, uses $\lg(1+\sqrt{2})H(\R) + 
O(t+n+\sigma)$ bits of space and supports the operations within the
complexities given in Table \ref{tab:time_wt_vs_brwt}.
\label{thm:brwt}
\end{theorem}

\begin{proof}
The operations have been obtained throughout the section.
For the space, $B$ is of length $\sigma+t$. Thus $O(t+n+\sigma)$ bits
account for $B$ and for the $2n$ bits at the root of the wavelet
tree. The rest of the bits in the wavelet tree can be counted by considering
that each bit not in the root is induced by the presence of a pair.

Each pair has a 
unique representative bit in a leaf, and also induces the presence of bits
up to the root. Yet those leaf-to-root paths get merged, so that not all those
bits are different. Consider an element $x$ related to $t_x$ labels. It induces
$t_x$ bits at $t_x$ leaves, and each such bit at a leaf induces a bit per
level on a path from the leaf towards the single $x$ at the root.%
\footnote{For example, in Figure~\ref{fig:br}, object $x=4$ is related to 
labels C and D (see also Figure~\ref{fig:binrel_example}). Its 1 at the second
leaf, for C, induces a 1 at its parent, for C-D, and a 1 at the root, for A-D. 
Its 1 at the third leaf, for E, induces a 1 at its parent for E-F and the 1 at 
the root for E-H. The fact that $(4,\mathrm{C})\in\R$ induces the creation of 
one column at the leaf for C and one at its parent. On the other hand,
there are two pairs related to object 1, but they are merged at the second
level and thus there is only one path arriving at the root.}
At worst, all the $O(t_x)$ bits up to level $\lg t_x$ are
created for these elements, and from there on all the $t_x$ paths are 
different, adding up a total of $O(t_x)+t_x \lg \frac{\sigma}{t_x}$ bits. 
Adding over all $x$ we get $O(t)+\sum_x t_x \lg\frac{\sigma}{t_x}$. This
is maximized when $t_x = t/n$ for all $x$, yielding 
$O(t) + t \lg\frac{\sigma n}{t} = H(\R) + O(t)$ bits.

Instead of representing two bitmaps (which would multiply the above value by 2),
we can represent a single sequence $B_v$ with
the possible values of the two bits at each position, 00, 01, 10, 11. Only at
the root is 00 possible. Except for those $2n$ bits, we can represent the
sequence over an alphabet of size 3 with a zero-order representation
\cite{GRR08}, to achieve at worst $(\lg 3)H(\R)+o(t)$ bits for this 
part while retaining constant-time $\rank$ and $\select$ over each $B_v^l$ and
$B_v^r$. (To achieve this, we maintain the directories
for the original bitmaps, of sublinear size.)

To improve the constant $\lg 3$ to $\lg (1+\sqrt{2})$, we consider that the
zero-order representation actually achieves $|B_v|H_0(B_v)$ bits.
We call $B_x$ the concatenation of all the symbols induced by $x$, 
$\ell_x = |B_x| \le t_x$, and $H_x = |B_x|H_0(B_x)$.
Assume the $t_x$ bits are partitioned into $t_{01}$ 01's, 
$t_{10}$ 10's, and $t_{11}$ 11's, so that $t_x = t_{01} + t_{10} + 2t_{11}$,
$\ell_x = t_{01} + t_{10} + t_{11}$, and 
$H_x = t_{01}\lg\frac{\ell_x}{t_{01}} + t_{10}\lg\frac{\ell_x}{t_{10}} +
       t_{11}\lg\frac{\ell_x}{t_{11}}$. As $t_{11} = (t_x-t_{01}-t_{10})/2$,
the maximum of $H_x$ as a function of $t_{01}$ and $t_{10}$ yields the worst
case at $t_{01} = t_{10} = \frac{\sqrt{2}}{4}t_x$, so $t_{11} =
(\frac{1}{2}-\frac{\sqrt{2}}{4})t_x$ and $\ell_x =
(\frac{1}{2}+\frac{\sqrt{2}}{4})t_x$, where $H_x = \lg(1+\sqrt{2})t_x$ bits. 
This can be achieved separately for each symbol. Using the same distribution of
01's, 10's, and 11's for all $x$ we add up to
$\lg(1+\sqrt{2})t \lg\frac{\sigma n}{t} + O(t) = \lg(1+\sqrt{2})H(\R)+O(t)$ bits.
(Note that, if we concatenate all the wavelet tree levels, the $H_x$ strings
are interleaved in this concatenation.)
\end{proof}

Note that this is a factor of $\lg(1+\sqrt{2}) \approx 1.272$ away of the 
entropy of $\R$. 

\begin{table}[tb]
\begin{center}
{\scriptsize 
\begin{tabular}{| l | p{3.6cm} | p{3.6cm} |} \hline
Operation & BRWT & {\sc BinRel-WT}\\
\hline\hline
%
%
$\relnum(\alpha,\beta,x,y)$ 
& $O(\beta-\alpha+\lg\sigma)$
& $O(\lg\sigma)$ \\ \hline

$\relrnk(\alpha,x)$ 
& 
  $O(\alpha+\lg\sigma)$
& $O(\lg\sigma)$ \\ \hline

$\relrnklabfst(\alpha,x,y,z)$ 
& $O(\alpha+\lg\sigma)$
& $O(\lg\sigma)$\\ \hline

$\relsellabfst(\alpha,j,x,y)$ 
& $O(j\lg\sigma)$
& $O(\lg\sigma)$\\ \hline

$\relminlabfst(\alpha,x,y,z)$ 
& $O(\lg\sigma)$
& $O(\lg\sigma)$\\ \hline

$\relrnkobjfst(\alpha,\beta,\gamma,x)$ 
&
  $O(\beta-\alpha+\lg\sigma)$
& $O(\lg\sigma)$\\ \hline

$\relselobjfst(\alpha,\beta,x,j)$ 
&  
  $O(j\lg\sigma)$
& 
  $O(\lg j\lg(\beta-\alpha+1)\lg\sigma)$

  $O(\lg n \lg\sigma)$ \\ \hline

$\relacc(\alpha,\beta,x,y)$ 
& $O((k+1)\lg\sigma)$
& $O((k+1)\lg\sigma)$\\ \hline

$\relminobjfst(\alpha,\beta,\gamma,x)$ 
& $O(\lg\sigma)$
& $O(\lg\sigma)$\\ \hline

$\labnum(\alpha,\beta,x,y)$ 
& $O(\beta-\alpha+\lg\sigma)$
& $O(\beta-\alpha+\lg\sigma)$ \\ \hline

$\labrnk(\alpha,x,y)$ 
& $O(\alpha+\lg\sigma)$
& $O(\alpha+\lg\sigma)$\\ \hline

$\labsel(\alpha,j,x,y)$ 
& $O(j\lg\sigma)$
& $O(j\lg\sigma)$\\ \hline

$\labacc(\alpha,x,y)$ 
& $O((k+1)\lg\sigma)$
& $O((k+1)\lg\sigma)$\\ \hline

$\labmin(\alpha,x,y)$ 
& $O(\lg\sigma)$
& $O(\lg\sigma)$\\ \hline

$\objnum(\alpha,\beta,x,y)$ 
& $O((y-x+1)\lg\sigma)$
& $O((y-x+1)\lg\sigma)$\\ \hline

$\objrnk(\alpha,\beta,x)$ 
& $O((y-x+1)\lg\sigma)$
& $O((y-x+1)\lg\sigma)$\\ \hline

$\objsel(\alpha,\beta,x,j)$ 
& $O(j\lg\sigma)$
& $O(j\lg\sigma)$\\ \hline

$\objacc(\alpha,\beta,x)$ 
& $O((k+1)\lg\sigma)$
& $O((k+1)\lg\sigma)$\\ \hline

$\objmin(\alpha,\beta,x)$ 
& $O(\lg\sigma)$
& $O(\lg\sigma)$\\ \hline

$\labrnko(\alpha,x)$ 
& $O(\lg\sigma)$
& $O(\lg\sigma)$\\ \hline

$\labselo(\alpha,j,x)$ 
& $O(j\lg\sigma)$
& $O(\lg\sigma)$\\ \hline

$\labmino(\alpha,j,x)$ 
& $O(\lg\sigma)$
& $O(\lg\sigma)$\\ \hline

$\labacco(\alpha,j,x)$ 
& $O((k+1)\lg\sigma)$
& $O((k+1)\lg\sigma)$\\ \hline

$\objrnko(\alpha,x)$ & $O(\lg\sigma)$ & $O(\lg\sigma)$\\ \hline
$\objselo(\alpha,x,j)$ & $O(\lg\sigma)$  & $O(\lg\sigma)$ \\ \hline
$\objmino(\alpha,x,j)$ & $O(\lg\sigma)$  & $O(\lg\sigma)$ \\ \hline
$\objacco(\alpha,x,j)$ & $O((k+1)\lg\sigma)$  & $O((k+1)\lg\sigma)$ \\ \hline
\end{tabular}
%
}\caption{Time complexity for the operations for
BRWT and {\sc BinRel-WT}. The parameter $k$ represents the size
of the output for the $\access$ operators; one can consider $k=1$ for the reductions shown in
Theorem \ref{thm:reductions}.}
\label{tab:time_wt_vs_brwt}
\end{center}
\end{table}

\section{Conclusions and Future Work}

Motivated by the many applications where a binary relation $\R$ between $\sigma$
labels and $n$ objects arises, we have proposed a rich set of primitives of 
interest in such applications. We first extended
existing representations and showed that their potential is very limited
outside single-row or single-column operations. Then we proposed a
representation called {\sc BinRel-WT}, that uses a wavelet tree to solve a
large number of operations in time $O(\lg\sigma)$. This structure has been 
already of use in particular cases, but here we have made systematic use of it
and exposed its full potential. Furthermore, we have extended the results to
generalized wavelet trees, to obtain the structure {\sc BinRel-GWT}. This
structure achieves $O(\lg\sigma/\lg\lg n)$ time for many operations. It had already
been used for range counting and reporting \cite{BHMM09}, but here we have 
extended its functionality to many other operations, so that its use as a 
replacement of the better known {\sc BinRel-WT} structure improves the times achieved in 
various applications by a $\lg\lg n$ factor.

Some of those speedups have already been mentioned in the article, namely
operations {\em range\_next\_value} and {\em prevLess} \cite{GNP11,KN11}
({\em range\_quantile}, on the other hand, is an interesting operation that
has resisted our attempts to improve it; it is equivalent to $\relsellabfst$).
Another example is finding the {\em dominant} points on a grid, that is, those
$(x,y)$ so that there is not another $(x',y')$ with $x'\le x$ and $y'\le y$. 
Using successive calls to $\relminobjfst$ and restricting the next call to the
area $[x+1,n] \times [1,y-1]$ where $(x,y)$ is the last pair found, structure 
{\sc BinRel-GWT} can find the dominant points in time $O(\lg n / \lg\lg n)$
per point, on an $n \times n$ grid. This improves upon the $O(\lg n)$ time
achieved in the literature \cite{NR11}.

The speedup on counting and reporting on grids is 
useful in various text indexing data structures. As an example, Arroyuelo et 
al.~\cite{ANS12} describe a Lempel-Ziv compressed index that is able 
to find the $occ$ occurrences of a pattern of length $m$ in a text $T[1,n]$
in time $O(m^2+(m+occ)\lg n)$. For this sake they use a 2-dimensional grid
where $O(m)$ $\relrnk$ and $O(occ)$ $\relacc$ operations are carried out. By
using our new {\sc BinRel-GWT} structure, their time complexity drops to
$O(m^2 + (m+occ)\lg n / \lg\lg n)$. As a second example, Claude and Navarro
\cite{CNfi10} use grammar-based compression to solve the same problem. 
Given a grammar of $n$ rules and height $h$, they achieve search time
$O(m(m+h)+h\,occ)\lg n)$. They use a 2-dimensional grid where operations 
$\relacc$, $\labselo$, and $\objselo$ are carried out, and therefore using 
a {\sc BinRel-GWT} structure their time is reduced to $O(m(m+h)+h\,occ)\lg n/\lg\lg n)$.

Despite the fact that our structures solve many of the queries we have proposed in
polylogarithmic (and usually logarithmic) time, 
supporting others remains a challenge. In particular, we have no good solutions
for label-level or object-level $\rank$ and $\select$ queries on ranges. 

Another challenge is space. All of the described structures use essentially 
$t\lg\sigma$ bits of space, where $t$ is the number of pairs in the relation.
While this is reasonable in many cases, it can be far from the entropy of the 
binary relation, $H(\R)$, for dense relations. We have proposed a variant, 
BRWT, that uses space within a multiplicative factor 
1.272 of $H(\R)$, yet its functionality is more limited: Apart from
label-level and object-level queries, this structure does not support to
efficiently count and select arbitrary points in ranges, albeit it can 
efficiently enumerate them. 

Since the publication of
the conference version of this article \cite{BCN10}, a followup work
\cite{FGN10} used our representation {\sc BinStr-WT} as an internal structure 
to achieve asymptotically optimum space, $H(\R) + o(H(\R))$ bits, and was
able to solve query $\relrnk$ in time $O(\lg n)$, and queries $\relacc$, 
$\relsellabfst$ and $\relselobjfst$ in time $O(\lg^2 n)$, on $n\times n$
grids. Once again, using structure {\sc BinRel-GWT} their times for $\relrnk$
and $\relacc$ can be divided by $\lg\lg n$.

It is therefore an open challenge to approach space $H(\R)$ as much as
possible while retaining the maximum possible functionality and the best
possible efficiency. It is unclear which are the limits in this space/time
tradeoff, although some lower bounds from computational geometry are useful.
For example, we cannot count points in ranges faster than the {\sc BinRel-GWT}
structure within polylogarithmic space \cite{Pat07}. 

On the other hand, $H(\R)$ is a crude measure that does not account
for regularities in the row/column distribution, or clustering, that arises in
real-life binary relations. For example, structures {\sc BinRel-WT} and
{\sc BinRel-GWT} can be compressed to the zero-order entropy of a string of
length $t$, and this can in some cases be below $H(\R)$, which shows that this
measure is not sufficiently refined. It would be interesting to consider finer 
measures of entropy and try to match them. 

There might also be other operations
of interest apart from the set we have identified.
For example, determining whether a pair is related in the {\em transitive
closure} of $\R$ is relevant for many applications (e.g., ancestorship
in trees, or paths in graphs). Another extension is to $d$-ary relations, which
would more naturally capture joins in the relational model.

Finally, we have considered only static relations. Our representations do allow
dynamism, where new pairs and/or objects can be inserted in/deleted 
from the waveleet trees \cite{MN08}. Adding/removing labels, instead, is 
an open challenge, as it alters the shape of the wavelet tree.  

\appendix

\section{Formal Definition of Operations}
\label{sec:defop}

\begin{figure}[p]
\begin{center}
\includegraphics[scale=0.6]{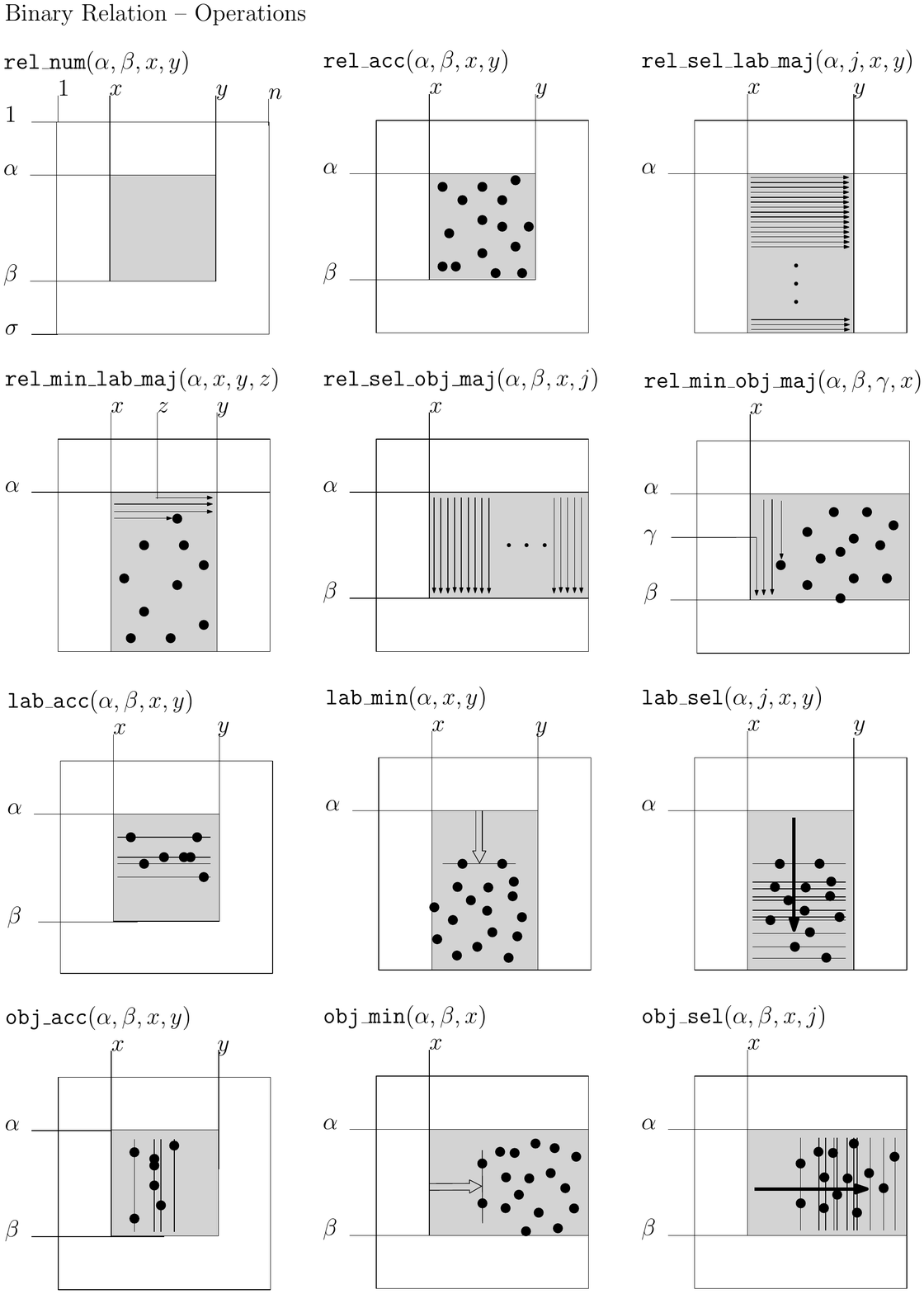}
\caption{Some operations illustrated.}
\label{fig:ops1}
\end{center}
\end{figure}

We formally define our set of operations. Figure \ref{fig:ops1} graphically
illustrates some of them.

\begin{itemize}
\item $\relacc(\alpha,\beta,x,y) ~=~ 
	\{ (\gamma,z) \in \R,\ \gamma \in [\alpha,\beta] ~\land~ z\in[x,y]\}$
\item $\relsellabfst(\alpha,j,x,y) ~=~	
	j\textrm{-th smallest pair of}~\relacc(\alpha,\sigma,x,y)$ 
	in order $(\alpha,x) \le (\beta,y) \Leftrightarrow
		\alpha < \beta \lor (\alpha=\beta \land x \le y)$
\item $\relminlabfst(\alpha,x,y,z) ~=~ 
	\textrm{under the same order, smallest pair of}$ $
	\relacc(\alpha,\alpha,z,y)\cup\relacc(\alpha{+}1,\sigma,x,y)$
\item $\relselobjfst(\alpha,\beta,x,j) ~=~	
	j\textrm{-th smallest pair of}~\relacc(\alpha,\beta,x,n)$ 
	in order $(\alpha,x) \le (\beta,y) \Leftrightarrow
		x < y \lor (x=y \land \alpha \le \beta)$
\item $\relminobjfst(\alpha,\beta,\gamma,x) ~=~ 
	\textrm{under the same order, smallest pair of}$ $
	\relacc(\gamma,\beta,x,x)\cup\relacc(\alpha,\beta,x{+}1,n)$
\item $\labacc(\alpha,\beta,x,y) ~=~ 
	\{ \gamma,\ \exists z,\ (\gamma,z) \in \relacc(\alpha,\beta,x,y) \}$
\item $\labacco(\alpha,\beta,x) ~=~ \labacc(\alpha,\beta,x,x)$
\item $\labsel(\alpha,j,x,y) ~=~ 
	j\textrm{-th smallest label of}~\labacc(\alpha,\sigma,x,y)$
\item $\labselo(\alpha,j,x) ~=~ \labsel(\alpha,j,x,x)$, 
	or $\relsellabfst(\alpha,j,x,x)$
\item $\labmin(\alpha,x,y) ~=~ \labsel(\alpha,1,x,y)$
\item $\labmino(\alpha,x) ~=~ \labmin(\alpha,x,x)$, or $\labselo(\alpha,1,x)$
\item $\objacc(\alpha,\beta,x,y) ~=~ 
	\{ z,\ \exists \gamma,\ (\gamma,z) \in\relacc(\alpha,\beta,x,y) \}$
\item $\objacco(\alpha,x,y) ~=~ \objacc(\alpha,\alpha,x,y)$
\item $\objsel(\alpha,\beta,x,j) ~=~ 
	j\textrm{-th smallest object of}~\objacc(\alpha,\beta,x,n)$
\item $\objselo(\alpha,x,j) ~=~ \objsel(\alpha,\alpha,x,j)$, 
	or $\relselobjfst(\alpha,\alpha,x,j)$
\item $\objmin(\alpha,\beta,x) ~=~ \objsel(\alpha,\beta,x,1)$
\item $\objmino(\alpha,x) ~=~ \objmin(\alpha,\alpha,x)$, or $\objselo(\alpha,x,1)$
\item $\relnum(\alpha,\beta,x,y) ~=~ |\relacc(\alpha,\beta,x,y)|$
\item $\relrnk(\alpha,x) ~=~ \relnum(1,\alpha,1,x)$
\item $\relrnklabfst(\alpha,x,y,z) ~=~ 
	\relnum(1,\alpha{-}1,x,y) {+} \relnum(\alpha,\alpha,x,z)$
\item $\relrnkobjfst(\alpha,\beta,\gamma,x) ~=~ 
	\relnum(\alpha,\beta,1,x{-}1) {+} \relnum(\alpha,\gamma,x,x)$
\item $\labnum(\alpha,\beta,x,y) ~=~ |\labacc(\alpha,\beta,x,y)|$
\item $\labrnk(\alpha,x,y) ~=~ \labnum(1,\alpha,x,y)$
\item $\labrnko(\alpha,x) ~=~ \labrnk(\alpha,x,x)$, or $\relnum(1,\alpha,x,x)$
\item $\objnum(\alpha,\beta,x,y) ~=~ |\objacc(\alpha,\beta,x,y)|$
\item $\objrnk(\alpha,\beta,x) ~=~ \objnum(\alpha,\beta,1,x)$
\item $\objrnko(\alpha,x) ~=~ \objrnk(\alpha,\alpha,x)$, or
	$\relnum(\alpha,\alpha,1,x)$
\end{itemize}

\section{Algorithms on {\sc BinRel-Str}}
\label{sec:binrelstr-proofs}

We prove some of the complexities in Table~\ref{tab:time_wt_vs_str}. The 
others can be derived using Theorem~\ref{thm:reductions}.

\begin{lemma}
\label{lem:brstrop1}
{\sc BinRel-Str} supports $\relnum(\alpha,\beta,x,y)$ in
$O(\min((\beta-\alpha+1)r, (y-x+1)a\lg\beta))$ time.
\end{lemma}

\begin{proof}
We can compute $\relnum(\alpha,\beta,x,y)$ in two ways:
\begin{itemize}
\item Using that $\relnum(\alpha,\beta,x,y) = \sum_{\alpha\le\gamma\le\beta}
\relnum(\gamma,\gamma,x,y)$ and that 
$\relnum(\gamma,\gamma,x,y) = \rank_\gamma(S,\map(x-1)+1,\map(y))$,
we achieve time $O((\beta-\alpha+1)r)$.
\item Using that $\relnum(\alpha,\beta,x,y) = \sum_{x\le z \le y}
\relnum(\alpha,\beta,z,z)$, we can compute the value for each $z$ by
searching for the successor of $\alpha$ and the predecessor of $\beta$ in
$S[\map(z-1)+1,\map(z)]$. As this range of $S$ is sorted we can find the
predecessors and successors using exponential search, which requires
in $O(\lg\beta)$ $\access$ operations. Thus the overall process takes
time $O((y-x+1)a\lg\beta)$.
\end{itemize}
\end{proof}



\begin{lemma}
\label{lem:brstrop2}
{\sc BinRel-Str} supports $\labnum(\alpha,\beta,x,y)$ in
$O(\min((\beta-\alpha+1)r, (y-x+1)(\lg\alpha+\beta-\alpha)a))$ time.
\end{lemma}

\begin{proof}
We can compute $\labnum(\alpha,\beta,x,y)$ in two ways:
\begin{itemize}
\item Using that $\labnum(\alpha,\beta,x,y) = \sum_{\alpha\le\gamma\le\beta}
\labnum(\gamma,\gamma,x,y)$, and that 
$\labnum(\gamma,\gamma,x,y) = 1$ iff $\relnum(\gamma,\gamma,x,y)>0$ and
zero otherwise, we can achieve the same time as in the first alternative
of Lemma~\ref{lem:brstrop1}.
\item Using that $\labnum(\alpha,\beta,x,y) = |\cup_{x \le z \le y}
\labacc(\alpha,\beta,z,z)|$, we can collect the labels in $[\alpha,\beta]\times
[z,z]$ for each $z$ and insert them into a dictionary. The labels related to 
a single $z$ can be found by using a similar method as the one in 
Lemma \ref{lem:brstrop1}: use exponential search to find the first element
$\ge \alpha$ in $z$'s area of $S$, and then scan the next symbols until
surpassing $\beta$. We mark each label found in a bitmap of length 
$\beta-\alpha+1$, and then we report the number of ones in it.
\end{itemize}
\end{proof}
\begin{lemma}

\label{lem:brstrop2.5}
{\sc BinRel-Str} supports $\objnum(\alpha,\beta,x,y)$ in
$O(\min((y-x+1)a\lg\alpha, (\beta-\alpha+1)(r+(y-x+1)s)))$ time.
\end{lemma}

\begin{proof}
We can compute $\objnum(\alpha,\beta,x,y)$ in two ways:
\begin{itemize}
\item Using that $\objnum(\alpha,\beta,x,y) = \sum_{x \le z \le y}
\objnum(\alpha,\beta,z,z)$, and that 
$\objnum(\alpha,\beta,z,z) = 1$ iff $\relnum(\alpha,\beta,z,z)>0$ and
zero otherwise, we can proceed similarly as in the second alternative
of Lemma~\ref{lem:brstrop1}, by exponentially searching for the first value
$\ge \alpha$ in $z$'s area of $S$ and checking whether it is $\le \beta$.
\item Using that $\objnum(\alpha,\beta,x,y) = |\cup_{\alpha\le\gamma\le\beta}
\objacc(\gamma,\gamma,x,y)|$, we can collect the objects in $[\gamma,\gamma]
\times [x,y]$ for each $\gamma$ and insert them into a dictionary, as in
Lemma~\ref{lem:brstrop2}. The objects related to a single $\gamma$ can be 
found by using successive $\select_\gamma(S,j)$ operations on
$S[\map(x-1)+1,\map(y)]$, starting with $j=\rank_\gamma(S,\map(x-1))+1$.
The complexity considers the worst case where each such $\gamma$ appears
$y-x+1$ times.
\end{itemize}
\end{proof}

Note that, when reducing to implement $\objrnk$, the $\rank_\gamma(S,\cdot)$
operation is not necessary. For $\objrnk1$ it is better to reduce from
$\relnum$.


\begin{lemma}
\label{lem:brstrop3}
{\sc BinRel-Str} supports $\relsellabfst(\alpha,j,x,y)$ in
$O(\min((\sigma-\alpha)r+s,(y-x+1)(\sigma-\alpha+1)a))$ time.
\end{lemma}

\begin{proof}
Again, we have two possible solutions:
\begin{itemize}
\item Set $c\leftarrow 0$. For each label $\gamma$ in $[\alpha,\sigma]$,
compute $c' \leftarrow c + \rank_\gamma(S,\map(x-1)+1,\map(y))$. If at some
step it holds $c' \ge j$, the answer is
 $(\gamma,\unmap(\select_\gamma(S,j-c+\rank_\gamma(S,\map(x-1))))$. 
Otherwise, update $c\leftarrow c'$. The overall process takes
$O((\sigma-\alpha+1)r+s)$ time.
\item Similarly, but first accumulating all the occurrences of all the labels
$\gamma$ and then finding $j$ using the accumulators. We simply traverse the
area $S[\map(z-1)+1,\map(z)]$ backwards, for each $z \in [x,y]$, accessing
each label $S[k]$ and incrementing the corresponding counter. The process takes
$O((y-x+1)(\sigma-\alpha+1)a)$ time.
\end{itemize}
\end{proof}

From this operation we can obtain complexities for 
$\relminlabfst$, $\labmin$, $\labmino$, $\labacc$, and $\labacco$.
Some of the results we give are better than those obtained by a blind 
reduction; we leave to the reader to check these improvements. We also
note that an obvious variant of this algorithm is our best solution to
compute $\labsel$, within the same time.

\begin{lemma}
\label{lem:brstrop4}
{\sc BinRel-Str} supports $\relselobjfst(\alpha,\beta,x,j)$ in
$O(\min((n-x+1)a\lg\beta,(\beta-\alpha+1)((n-x+1)s+r)))$ time.
\end{lemma}

\begin{proof}
Once again, we have two possible solutions:
\begin{itemize}
\item Set $c\leftarrow 0$. For each object $z$ in $[x,n]$,
use exponential search on $z$'s area of $S$ to find the range 
$S[a,b]$ corresponding to $[\alpha,\beta]$, and set $c' \leftarrow c + b-a+1$.
If at some step it holds $c' \ge j$, the answer is $(S[j-c+a-1],z)$.
Otherwise, update $c\leftarrow c'$. The overall process takes
$O((n-x+1)a\lg\beta)$ time.
\item Similarly, but first accumulating all the occurrences of all the objects
$z$ and then finding $j$ using the accumulators. We traverse the
area $S[\map(x-1)+1,\map(y)]$ for each label $\gamma$ in $[\alpha,\beta]$, 
using successive $\select_\gamma(S,j')$ queries, starting at 
$j'=\rank_\gamma(S,\map(x-1))+1$. The process takes 
$O((\beta-\alpha+1)((n-x+1)s+r))$ time.
\end{itemize}
\end{proof}

From this operation we can obtain complexities for 
$\relminobjfst$, $\relacc$, $\objmin$, $\objmino$, $\objacc$, and $\objacco$.
Once again, some of the results we give are better than a blind reduction and
we leave the reader to verify those. Finally, an obvious variant of this
algorithm is our best solution to compute $\objsel$.

%

A final easy exercise for the reader is to show that $\labselo(\alpha,j,x)$
can be solved in time $O(a\lg\alpha)$, and that $\objselo(\alpha,x,j)$ is
solved in time $O(r+s)$.

\bibliographystyle{abbrv}
\bibliography{paper}

\end{document}